\documentclass[12pt,nofootinbib]{revtex4-1}

\usepackage[utf8]{inputenc}
\usepackage{amsmath}
\usepackage{amssymb}
\usepackage{amsfonts}
\usepackage{graphicx}
\usepackage{color}
\usepackage{amsbsy}
\usepackage{url}
\usepackage{enumerate}
\usepackage[vcentermath,enableskew]{youngtab}
\usepackage{verbatim}
\usepackage[normalem]{ulem}
\usepackage{cancel}
\usepackage{bm,bbm}
\usepackage{slashed}

\makeatletter
\@addtoreset{equation}{section}

\makeatother

\newcommand{\refer}[1]{(\ref{#1})}

\newcommand{\abs}[1]{\left|#1\right|}

\newcommand{\sech}{\mbox{sech}}

\def\be{\begin{eqnarray}}
\def\ee{\end{eqnarray}}
\def\p{\partial}

\begin{document}
\vspace*{-2cm}
\begin{flushright}
YGHP-18-02
\end{flushright}

\title{
Localized non-Abelian gauge fields in non-compact extra-dimensions
}

\author{Masato Arai$^1$, Filip Blaschke$^{2, 3}$, Minoru Eto$^4$ and Norisuke Sakai$^5$\ \\\ }
\affiliation{
$^1$Faculty of Science, Yamagata University, 
Kojirakawa-machi 1-4-12, Yamagata,
Yamagata 990-8560, Japan\\
$^2$Faculty of Philosophy and Science, Silesian University in Opava, Bezru\v{c}ovo n\'am. 1150/13, 746~01 Opava, Czech Republic\\
$^3$Institute of Experimental and Applied Physics, Czech Technical University in Prague, Horsk\'a 3a/22, 128 00 Praha 2, Czech Republic\\
$^4$Department of Physics, Yamagata University, 
Kojirakawa-machi 1-4-12, Yamagata,
Yamagata 990-8560, Japan\\
$^5$Department of Physics, and Research and 
Education Center for Natural Sciences, 
Keio University, 4-1-1 Hiyoshi, Yokohama, Kanagawa 223-8521, Japan\\
and iTHEMS, RIKEN,
2-1 Hirasawa, Wako, Saitama 351-0198, Japan
}

\begin{abstract}
\ \\
Dynamical localization of non-Abelian gauge fields in non-compact 
flat $D$ dimensions is worked out. 
The localization takes place via a field-dependent 
gauge kinetic term when a field  
condenses in a finite region of  
spacetime.
Such a situation typically arises in the presence of 
topological solitons.  
We construct  four-dimensional low-energy effective Lagrangian 
up to the quadratic order in a universal manner 
applicable to any spacetime dimensions. 
We devise an extension of the $R_\xi$ gauge 
to separate physical and unphysical modes clearly. 
Out of the D-dimensional non-Abelian gauge fields, the physical 
massless modes reside only in the four-dimensional components,  
whereas they are absent in the extra-dimensional components.
The universality of non-Abelian gauge charges holds due to 
the unbroken four-dimensional gauge invariance.  
We illustrate our methods with models in $D=5$ (domain walls), 
 in $D=6$ (vortices), and in $D=7$.
\end{abstract}

\maketitle

\newpage

%


\section{Introduction and Conclusions}

Theories with extra-dimensions give a solution of the gauge 
hierarchy problem in the framework such as the brane-world 
scenario \cite{ArkaniHamed:1998rs,Antoniadis:1998ig, 
RS,RS2,Akama}.  One of the most popular models is in five-dimensional 
spacetime where the fifth dimension is {\it compactified} on an orbifold.
In this kind of models, several assumptions are made:
i) The fifth dimension is compact. 
ii)  Branes exist. iii) Matter fields are localized on the branes 
with the boundary Lagrangian (proportional to a delta function). 
iv) Nontrivial $\mathbb{Z}_2$ parity assignments are imposed 
on fields.
This setup provides models akin to the standard model 
(SM) with several nice solutions to long-standing problems of the SM. 
However, the origins of these nontrivial assumptions have not 
been explained.

In contrast, these points can be achieved not as assumptions 
but as consequences of dynamics in a model with {\it non-compact} 
extra-dimensions. 
We do not need to prepare a specific geometry for the extra-dimensions. 
For five-dimensional models, the minimal assumption is 
the presence of discrete degenerate vacua. 
Spontaneous symmetry breaking of the discrete symmetry dynamically 
yields stable domain walls. 
Thus, our four-dimensional world is dynamically realized on 
the domain walls. 
Furthermore, they automatically lead to localization of zero 
modes of matter fields such as chiral fermions 
and scalars, out of D-dimensional matter fields in the bulk
 \cite{Jackiw:1975fn,Jackiw:1981ee,Rubakov}.
The extra-dimensional models can give a natural 
explanation also for the hierarchy among the effective 
four-dimensional Yukawa couplings 
\cite{ArkaniHamed:1999dc,Dvali:2000ha,Gherghetta:2000qt,Kaplan:2000av,Huber:2000ie}, irrespective of compact or non-compact extra dimensions.

Unfortunately, the localization of gauge fields is quite 
difficult  \cite{Dvali:2000rx, Kehagias:2000au, Dubovsky:2001pe, 
Ghoroku:2001zu,Akhmedov:2001ny, Kogan:2001wp, Abe:2002rj, 
Laine:2002rh, Maru:2003mx, Batell:2006dp, Guerrero:2009ac, 
Cruz:2010zz, Chumbes:2011zt, Germani:2011cv, Delsate:2011aa, 
Cruz:2012kd, Herrera-Aguilar:2014oua, Zhao:2014gka, Vaquera-Araujo:2014tia,
Alencar:2014moa,Alencar:2015awa,Alencar:2015rtc,Alencar:2015oka,Alencar:2017dqb} 
in the brane-world scenario with the topological defects.
A popular resolution is the so-called Dvali-Shifman mechanism 
\cite{Dvali:1996xe}. 
However, this mechanism assumes the confinement in 
higher-dimensional spacetime, whose validity is far from being clear. 
It was found that gravity can localize gauge fields but it works 
only for six dimensions \cite{Oda:2000zc}.
A problem of using gravity is that gravity affects all the fields 
on an equal footing.
While the gauge fields localization may be achieved, the localization 
of fermions is lost \cite{Oda:2000zc}. 

It has been noted that the localization of gauge fields in 
flat non-compact spacetime requires the confining phase rather 
than the Higgs phase in the bulk outside the brane 
\cite{Dvali:1996xe,ArkaniHamed:1998rs}. 
A semi-classical realization of the confinement can be obtained by 
the position-dependent gauge coupling \cite{Kogut:1974sn,Fukuda:1977wj, Fukuda:2009zz, Fukuda:2008mz}, 
which is achieved by domain walls in five dimensions through the 
field-dependent gauge coupling function. This semi-classical 
mechanism was successfully applied to localize gauge fields 
on domain walls \cite{Ohta:2010fu,Arai:2012cx,Arai:2013mwa,Arai:2014hda,Arai:2016jij,Arai:2017lfv,Arai:2017ntb}. 
As an advantage of using this mechanism, we can explicitly 
determine mode functions of massless gauge bosons. 
Recently, we proposed a model realizing Grand Unified Theory (GUT) 
on  domain walls in five dimensions \cite{Arai:2017ntb} via 
the geometric Higgs mechanism \cite{Arai:2017lfv} which gives 
the familiar Higgs mechanism by means of the geometric information (position of walls along the extra-dimension).
Another advantage is that our localization mechanism assures 
charge universality of matter fields 
by preserving the $3+1$-dimensional gauge invariance.

The main goal of this paper is to establish a rigorous formulation 
of the localization mechanism of gauge fields by generalizing 
the non-trivial gauge kinetic function 
from five dimensions 
\cite{Ohta:2010fu,Arai:2012cx,Arai:2013mwa,Arai:2014hda,Arai:2016jij,Arai:2017lfv,Arai:2017ntb} to higher dimensions.
With this established formula at hand, one can naturally construct 
brane-world models in higher dimensions without assuming either 
a compact geometry or the confinement in higher dimensions. 
Especially, the models in six dimensions has a nice virtue that 
a single family in six dimensions automatically generates $k$ 
copies of massless fermions in four dimensions 
where $k \in \pi_1(S^1)$ is the topological vortex number, 
giving an explanation of the three generations in the SM. 
A similar mechanism has been discussed in models without the 
localization of gauge fields \cite{Libanov:2000uf,Frere:2000dc}, 
assuming Dvali-Shifman mechanism \cite{Frere:2001ug}, and 
with $S^2$ as the compact extra dimensions \cite{Frere:2003yv}.

In this work, we study the localization of non-Abelian gauge fields 
as generically as possible.
Our analysis  
is quite model independent and it   is 
applicable to  any number of  spacetime dimensions. 
Since our primary aim here is to clarify the physical mass 
spectrum appearing in low energy four-dimensional physics, 
we will analyze the action up to the quadratic order in fields. 
Then, we can treat Abelian and non-Abelian gauge fields on 
equal footing. 
Although we do not consider quantum loop calculations in this 
paper, we  develop an extension of $R_\xi$ gauge appropriate 
for models in higher dimensions in order to separate physical and unphysical
degrees of freedom. 
In  contrast, let us recall 
our previous studies in five dimensions 
\cite{Ohta:2010fu,Arai:2012cx,Arai:2013mwa,Arai:2014hda,Arai:2016jij,Arai:2017lfv,Arai:2017ntb} where 
the axial gauge $A_y = 0$ was chosen. 
Although the axial gauge is simple, it is inappropriate 
to establish the possible presence of zero modes of $A_y$ besides 
being awkward for loop calculations. 
One should note that the zero mode is gauge invariant.
The analysis in our $R_\xi$-like gauge will not only provide 
clearcut understanding of the physical spectrum but also is applicable 
to higher dimensions $D\ge 6$ where the axial gauge $A_y=0$ does not naively make sense. 
It is gratifying that we do not find any additional zero modes 
except for desired four-dimensional gauge fields 
in low-energy effective theory: a fact that is also insensitive 
to the details of the model. 
This is due to the fact that the field dependent-gauge coupling function spontaneously
breaks the gauge symmetry in such a way that the gauge symmetry only in the four dimensional sense
is preserved.
In comparison, the standard compactification of extra-dimensions 
cannot avoid new zero modes from extra components of the gauge fields, 
and an additional structure, such as orbifolding, is required to 
suppress them. 
This point offers a possibility for our mechanism to become a 
universal tool for the brane-world model building.

To be concrete, we give two examples: one is a domain wall in 
five dimensions and the other is a Nielsen-Olsen type local 
vortex in six dimensions. 
While so many works have been done to localize gauge fields on 
domain walls, the number of works are quite \emph{a} few on the vortices. 
In particular, if we do not assume the Dvali-Shifman mechanism 
\cite{Frere:2001ug}, compact extra dimensions \cite{Frere:2003yv}, 
or do not use gravity \cite{Oda:2000zc}, the example given here 
is the first model which provides massless non-Abelian gauge 
fields on the vortices in six dimensions. 
We also give an example for $D=7$ case. 
We emphasize that our localization mechanism 
automatically gives universality of gauge charges in models 
in any dimensions \cite{Dvali:1996xe,Rubakov:2001kp}.

To analyze the physical spectra  
in non-compact spacetime, we find a formulation similar to the 
supersymmetric quantum mechanics quite useful. 
When we determine mass spectra of Kaluza-Klein (KK) modes, we always end up 
with a  Schr\"odinger type problem. 
The corresponding Hamiltonians we will encounter 
are indeed special ones. 
In five-dimensional case, they are precisely the Hamiltonians 
of supersymmetric (SUSY) quantum mechanics (QM).
Therefore, the spectra can be analytically obtained in many cases. 
In the higher-dimensional cases with $D\ge6$, 
the Hamiltonians are still similar to SUSY QM ones. 
This structure is very helpful both analytically and numerically.

This paper is organized as follows. In Sec.~\ref{sec:2} we present 
a general argument of gauge field localization in general $D$-dimensions. 
Furthermore, we separate physical and unphysical modes of massive 
as well as massless four-dimensional fields, and work out the 
low-energy effective theory. 
We extend the $R_\xi$ gauge and develop a SUSY QM technique. 
In Sec.~\ref{sec:3} we provide three explicit examples of 
brane-world scenarios with models of one non-compact extra 
dimensions (domain walls), two extra dimensions (vortices), and 
three extra dimensions.  

Note added: While finishing this work, a new paper \cite{Okada:2017omx}
appeared  that has a partial overlap with some of our results.
Just after this paper was posted on arXiv, another new paper \cite{Okada:2018von} appeared.

\section{Localization and Higss-like mechanisms}
\label{sec:2}

\subsection{Generic formula}

Let us consider a simple Yang-Mills model in $D$ dimensions
\be
{\cal L}_A = - \beta^2\, {\rm Tr}\,{\cal F}_{MN} {\cal F}^{MN},
\label{eq:lag}
\ee
where ${\cal F}_{MN} = \p_M {\cal A}_N - \p_N {\cal A}_M 
+ i [{\cal A}_M,{\cal A}_N]$ is a non-Abelian field strength. 
Throughout the paper, we use small greek letters for 
four-dimensional indices $\mu = 0,1,2,3$,
small roman letters for extra-dimensional spatial coordinates 
$a=4,\cdots,D-1$ and the capital roman letters for the 
$D$-dimensional indices $M=0,1,\cdots,D-1$. 
Mass dimensions of the gauge fields and $\beta$ are 
$[{\cal A}_M] = 1$ and $[\beta] = \frac{D-4}{2}$, respectively. 
We assume that $\beta$ is a Lorentz scalar and a gauge invariant.
We denote the four-dimensional coordinates as $x = \{x^\mu\}$ 
the extra-dimensional coordinates as $y=\{x^a\}$, and the metric as 
$\eta_{\mu\nu}=(1,-1,\cdots,-1)$. 
The non-minimal gauge kinetic term of type (\ref{eq:lag}) are studied in various contexts though
most of them concern four dimensions \cite{Cho:1986kd,Cho:1987yc,Cho:1992yj,Babichev:2006cy,Ramadhan:2015qku,Bazeia:2012uc,Atmaja:2015lia,Cho:2013vba,Ellis:2016glu,Blaschke:2017pym}.

When $\beta$ is a constant, it is nothing but the inverse gauge coupling constant, i.e. $\beta^{-1} = \sqrt{2}g$.
In this work, we will investigate what happens when $\beta$ is not a constant.
There are at least three cases where this situation is realized:
i) the spacetime geometry is nontrivial  \cite{Randall:2001gb} with 
$\sqrt{-g}$ identified as $\beta^2$, 
ii) $\beta^2$ is identified \cite{Kehagias:2000au} as $\beta^2 = e^{\varphi}$ 
with the dilaton field $\varphi$, 
iii) the gauge coupling is a function of scalar fields $\varphi_i$ 
as $\beta = \beta(\varphi_i)$ with $\varphi_i$ acquiring nonvanishing 
$y$-dependent vacuum expectation values inside a finite region 
in the extra-dimensions\footnote{
Thorough out this paper, we assume $\varphi_i$ to be singlet of the gauge group of ${\cal A}_M$. Therefore,
the condensation of $\varphi_i$ does not directly lead to spontaneous break down of the gauge symmetry.
The singlet scalar $\varphi_i$ interacts with ${\cal A}_M$ only through Eq.~(\ref{eq:lag}).}.
Each has its own (dis)advantages, but 
all the technical aspects, which we investigate here, are applicable for all of the cases.

Minimal assumption for us is that $\beta$ depends only on the 
extra-dimensional coordinates $y$. 
We further assume the square integrability 
\be
\int d^{D-4}x\, \beta(y)^2 < \infty.
\label{eq:sqint}
\ee
As stated above, the reason why $\beta$ depends on $y$ is not important for our results.
For concreteness, however, we will give several examples in later sections. 

The square integrability condition implies
that $\beta$ approaches zero as $|y| \to \infty$. This means that the gauge coupling become very large in the bulk.
This is a semiclassical realization of the confining phase in the bulk, which is necessary to realize localization of
the massless gauge fields on branes \cite{Dvali:1996xe,ArkaniHamed:1998rs}.

We first introduce differential operators which will play a central role in what follows:
\be
D_a = - \p_a + (\p_a \log \beta) = - \beta \p_a \frac{1}{\beta},\quad D_a^\dagger = \p_a + (\p_a \log\beta) = \frac{1}{\beta} \p_a \beta.
\ee
It is straightforward to verify the following
\be
D_a^\dagger D_b &=& - \frac{1}{\beta} \p_a\beta^2\p_b \frac{1}{\beta},\\
D_a D_b^\dagger &=& - \beta \p_a\frac{1}{\beta^2}\p_b \beta,\\ 
\left[D_a,D_b^\dagger\right] &=& -2\left(\p_a\p_b \log \beta\right),\label{eq:com}\\
\Bigl[D_a,D_b\Bigr] &=& \left[ D_a^\dagger, D_b^\dagger\right] = 0.
\ee
Throughout the paper, we will use the convention that the derivatives acts on everything to the right, unless explicitly delimited by parenthesis as shown in \refer{eq:com}.
Let us define an analog to a superpotential in one-dimensional 
SUSY 
quantum mechanics
\be
W_a = (\p_a \log \beta) = \frac{(\p_a \beta)}{\beta}.
\ee
Since $D_a = - \p_a + W_a$, $D_a^\dagger = \p_a + W_a$, 
we define\footnote{Here and in the following, we use a convention 
to sum over repeated indices unless stated  otherwise. } 
\be
D^2 \equiv D_a^\dagger D_a = - \p_a^2 + W_a^2 + (\p_a W_a) 
= - \p_a^2 + \frac{(\p_a^2 \beta)}{\beta},\\ 
\bar D^2 \equiv D_a D_a^\dagger = - \p_a^2 + W_a^2 - (\p_a W_a) 
= - \p_a^2 + \frac{(\p_a^2 \beta^{-1})}{\beta^{-1}},
\ee
 Let $\phi_n$ and $\bar\phi_n$ be eigenfunctions of $D^2$ and 
$\bar D^2$, respectively.
\be
D^2 \phi_n = m_n^2 \phi_n,\quad
\bar D^2 \bar \phi_n = \bar m_n^2 \bar \phi_n.
\label{eq:eigen_eq}
\ee
Here $n$ is symbolic index suitably labelling both discrete and continuum parts of the spectrum, including possible degenerate states. 
Note that $D^2$ and $\bar D^2$ are semi-positive definite operators, so that their eigenvalues are real and nonnegative.
We normalize the eigenfunctions by
\be
\int d^{D-4}x\, \phi_m\phi_n =  \delta_{mn},\quad
\int d^{D-4}x\, \bar\phi_m \bar\phi_n =  \delta_{mn},
\ee
where $\delta_{mn}$, again, symbolically represent both Kronecker's delta for discrete modes and delta function for continuum modes.
The mass dimension is $\left[\phi_n\right] = \left[\bar\phi_n\right] = \frac{D-4}{2}$.

Clearly, $D^2$ has a zero eigenfunction $\phi_0$, with eigenvalue $m_0 = 0$,
given as 
\begin{equation}
\phi_0 = N_0\beta\,.
\label{eq:zero}
\end{equation}
It's normalizability is ensured by  square-integrability of $\beta$. 
It will be proven that the zero eigenfunction $\phi_0 \propto \beta$ 
is important to assure the universality 
of non-Abelian gauge charges in four-dimensional effective Lagrangian.
Uniqueness of the normalizable zero eigenfunction can be easily 
shown at least for the case where $\beta = \beta(r)$ depends 
only on radial coordinate $r = \sqrt{x_a^2}$.  
Let us first note that $\phi_0$ should be a function of $r$ 
only because energy inevitably increases if $\phi_0$ depends 
on angular coordinates. 
Then let us rewrite Eq.~(\ref{eq:eigen_eq}) in terms of 
$\varphi_0 = r^{\frac{D-5}{2}} \phi_0$ as
\be
\left(- \frac{d^2}{dr^2} + {\cal V}(r)\right)\varphi_0 = 0,
\quad
{\cal V} = \frac{(\p_a^2 \beta)}{\beta} + \frac{(D-7)(D-5)}{4r^2}.
\ee
Square integrability condition is $\int  dr\, r^{D-5} \phi_0^2 
= \int dr\, \varphi_0^2 < \infty$.
Since this is nothing but a problem of one-dimensional 
quantum mechanics, all bound 
states are nondegenerate. 
Hence, the normalizable zero eigenfunction (\ref{eq:zero}) is unique at least in rotationally invariant backgrounds.

We also see that a solution to the equation $\bar D^2 \bar\phi_0 = 0$ is given as
\be
\bar\phi_0 \propto \beta^{-1}.
\ee
However, this is not square-integrable and, hence, not a part of a physical spectrum. 

Fundamental mass scales involved in
the Schr\"odinger problem are
\be
\Omega_a =\lim_{|y|\to\infty} \abs{\p_a \log \beta}.
\ee
We assume that there is a mass gap of order $\Omega_a$ between the zero mode $m_0 = 0$ 
and massive modes $m_n$ ($n\neq 0$).

Our primary interest is to work out physical spectra in the low-energy four-dimensional physics. Therefore 
we will consider the action up to the quadratic order in fields.  
As a consequence, non-Abelian gauge fields and Abelian gauge fields can be treated on the same footing.
For ease of notation, we will concentrate on the Abelian case in what follows.

In order to find physical degrees of freedom and mass spectra,  we have to find a suitable 
gauge fixing condition. 
Inspired by the usual $R_\xi$ gauge, we choose the 
gauge-fixing Lagrangian as 
\be
{\cal L}_{\rm gf} = - \frac{2\beta^2}{\xi} f^2,\quad
f = \p^\mu {\cal A}_\mu + \xi \frac{1}{\beta^2} \p^a \beta^2 {\cal A}_a,
\label{eq:gff}
\ee
where $\xi$ is an arbitrary constant. 
Note that if $\beta$ is a constant and take $\xi = 1$, this is 
nothing but the gauge fixing condition of the covariant
gauge $f = \p^M {\cal A}_M$ in $D$ dimensions.  
On the other hand, if we replace 
$\beta^{-2}\p^a \beta^2 {\cal A}_a$ by $m_h h$ 
as product of ``Higgs" mass $m_h$ and a ``Nambu-Goldstone field" $h$, 
it is almost identical to the gauge fixing functional 
used in the familiar $R_\xi$ gauge for the Higgs mechanism in four dimensions. 
The reason for the choice of $f$ in Eq.~(\ref{eq:gff}) is to eliminate mixing between the four-dimensional 
gauge fields ${\cal A}_\mu$ and extra-dimensional gauge fields ${\cal A}_a$.

Even though our analysis is essentially Abelian, it proves useful to investigate spectra of ghost fields as well, in order to clearly identify unphysical degrees of freedom. 
By varying the gauge fixing functional $f$ in Eq.~(\ref{eq:gff}), 
we find the ghost action as 
\be
{\cal L}_{\rm gh} = - \bar {\cal C} \left(\partial^2 + \xi \frac{1}{\beta^2}\p^a \beta^2 \p_a\right) {\cal C},
\ee
with the mass dimensions $[{\cal C}] = 1$ and $[\bar {\cal C}]=D-3$.

For further convenience, let us switch to the canonically normalized fields 
\be
{\cal A}_M = \frac{A_M}{2\beta},\quad {\cal C} = \frac{c}{\beta},\quad \bar {\cal C}= \beta \bar c.
\ee
Mass dimensions of these fields are given as $[A_M] = [c] = [\bar c] = \frac{D-2}{2}$.
In terms of the new fields, after performing integration by parts, 
the Lagrangians \refer{eq:lag} can be 
expressed as
\be
{\cal L}_A &=& \frac{1}{2} A_\mu \left(\eta^{\mu\nu}\p^2 - \p^\mu\p^\nu + \eta^{\mu\nu}D^2\right) A_\nu \nonumber\\
&&-\, \frac{1}{2} A_a \left(\delta_{ab} D^2 - D_b^\dagger D_a + \delta_{ab}\p^2 \right) A_a 
-  (D_a^\dagger A_a)  \p^\mu A_\mu,\\
{\cal L}_{\rm gf} &=& \frac{1}{2\xi}A_\mu\p^\mu\p^\nu A_\nu +  (D_a^\dagger A_a)\p^\mu A_\mu 
-\frac{1}{2}\xi A_a D_a D_b^\dagger A_b,\\
{\cal L}_{\rm gh} &=& - \bar c\left(\p^2 + \xi D^2\right) c,
\ee
with $\p^2 = \p_\mu \p^\mu$.
Collecting all pieces, we find our Lagrangian is of the form
\be
{\cal L}_\xi &=& \frac{1}{2} A_\mu \left[
\eta^{\mu\nu}\p^2 - \left(1-\frac{1}{\xi}\right)\p^\mu\p^\nu + \eta^{\mu\nu}D^2\right] A_\nu \nonumber\\
&&-\,\frac{1}{2} A_a\left[
\delta_{ab} D^2 - \left(D_b^\dagger D_a - \xi D_a D^\dagger_b\right) + \delta_{ab}\p^2\right] A_b \nonumber\\
&&-\, \bar c \left(\p^2 + \xi D^2\right)c.
\label{eq:lag_xi_full}
\ee
Interestingly, the four-dimensional part and the extra-dimensional part have similar structure
under the exchange of $\p_\mu$ and $D_a$.
The gauge fixing parameter $\xi$ serves as a mark to distinguish physical and unphysical degrees of freedom.

\subsection{Four-dimensional components of gauge fields $A_\mu$}

Firstly, let us investigate the first line of Eq.~(\ref{eq:lag_xi_full}).
It is quite similar to the  Lagrangian of the gauge theory in four dimensions.
The differences are that $A_\mu$ is function of not only $x=\{x^\mu\}$ but also $y=\{x^a\}$, and
$D^2$ is not a mass but the differential operator in terms of $\p_a$.

In order to get the physical spectrum, let us expand $A_\mu$ 
in terms of the eigenfunctions of $D^2$ defined
in Eq.~(\ref{eq:eigen_eq}) as
\be
A_\mu = A_\mu^{(0)}(x) \phi_0(y) + \sum_{n\neq0}A_\mu^{(n)}(x) \phi_n(y).
\ee
Since $[A_\mu] = \frac{D-2}{2}$ and $[\phi_n] = \frac{D-4}{2}$, this expansion ensures for
the four-dimensional gauge fields 
$A_\mu^{(n)}(x)$ to have 
correct mass dimension $[A_\mu^{(n)}]=1$.
Plugging this into the first line of Eq.~(\ref{eq:lag_xi_full}) and integrate it over the extra-dimensions, we get
\be
\int d^{D-4}x\, {\cal L}_\xi\big|_{\text{1st}} &=& 
\frac{1}{2} A_\mu^{(0)} \left[
\eta^{\mu\nu}\p^2 - \left(1-\frac{1}{\xi}\right)\p^\mu\p^\nu \right] A_\nu^{(0)}\nonumber\\
&&+\,\sum_{n\neq0}
\frac{1}{2} A_\mu^{(n)} \left[
\eta^{\mu\nu}\p^2 - \left(1-\frac{1}{\xi}\right)\p^\mu\p^\nu + \eta^{\mu\nu} m_n^2 \right] A_\nu^{(n)}.
\label{eq:L1st_4dim}
\ee

Note that in terms of the original field ${\cal A}_\mu$ the above expansion is rewritten as
\be
{\cal A}_\mu =\frac{N_0}{2} {\cal A}_\mu^{(0)}(x)  + \sum_{n\neq0}{\cal A}_\mu^{(n)}(x) \frac{\phi_n(y)}{2\beta}.
\ee
Remarkably, the zero mode wave function is constant in $y$. This ensures the universality of non-Abelian gauge charges
of matter fields, since overlap integral of the wave functions of gauge field and matter fields do not depend on the details of the localization mechanism \cite{Rubakov:2001kp,Arai:2017lfv,Arai:2017ntb}.

\subsection{Extra-dimensional components of gauge 
fields $A_a$}\label{sec:extrag}

Let us next investigate the physical spectrum of  
extra-dimensional gauge fields $A_a$
from the second line of Eq.~(\ref{eq:lag_xi_full}).
From the viewpoint of four dimensions, they are  
scalar fields.

We first consider the extra-dimensional divergence 
$K = D_a^\dagger A_a$. 
By applying $D_a^\dagger$ on the field equation for $A_a$, we 
obtain the field equation for $K$ as 
$
(\partial^2+\xi D^2)K=0. 
$
Hence we expand $K$ in terms of the eigenfunctions 
of $D^2$ as
\be
K(x,y) = -\Omega \phi_0(y) K^{(0)}(x) - \sum_{n\neq0} m_n \phi_n(y) K^{(n)}(x).
\label{eq:exp_div_6}
\ee
Note that the mass dimensions are $[K]=D/2$ and
 $[K^{(n)}]=1$ due to the intentional insertion of $\Omega$ and $m_n$. 
In the following, however, we will demonstrate the absence 
of zero mode $K^{(0)}(x)=0$. 
Let us suppose that there is a zero mode 
\begin{equation}
D_a^{\dagger}A_a = -\Omega \phi_0(y)K^{(0)}(x) \equiv k(x) \beta(y)\,.
\end{equation}
Multiplying this by $\beta$ we obtain
\begin{equation}
\partial_a\bigl(\beta A_a\bigr) = k \beta^2\,.
\end{equation}
Now we integrate this over extra-dimensions. 
The right-hand side is non-zero due to our square-integrability 
condition on $\beta$. However, for \emph{regular} $A_a$ the 
left-hand side is 
\begin{equation}
\int d^{D-4}x\, \partial_a\bigl(\beta A_a\bigr) 
= \int d^{D-3} S_a\, \bigl(\beta A_a\bigr) = 0\,,
\end{equation}
since $\beta$ vanishes at the boundary. 
We arrive at the contradiction, which shows the absence of 
zero mode: $K^{(0)}(x)=0$.

Absence of zero mode implies that $D^{-2}$ is well-defined 
on $K$. Hence we can define a projection operator $P_{ab}$ acting on 
$A_a$ to obtain the divergence part $A_a^{\rm d}$ 
\begin{equation}
A_a^{\rm d} = P_{ab} A_b = D_a D^{-2}K .
\label{eq:div_ex}
\end{equation}
\be
P_{ab} = D_a D^{-2} D_b^\dagger. 
\ee
The operator $P_{ab}$ enjoys the properties of a projection operator: 
\be
P_{ab}P_{bc} = P_{ac},\quad
\left(\delta_{ab}D^2 - D_b^\dagger D_a\right) P_{bc} = 0,\quad
\left(\delta_{ab}-P_{ab}\right)D_b D_c^\dagger = 0.
\ee
The remaining part is defined as divergence-free part: 
$A_a = A_a^{\rm d} + A_a^{\rm df}$ 
\be
A_a^{\rm df} = (\delta_{ab} - P_{ab})A_b,
\ee
These parts satisfy 
\be
D_a^\dagger A_a^{\rm d} = K,\qquad D_a^\dagger A_a^{\rm df} = 0.
\ee

By using the above identities, we can rewrite the second line of 
Eq.~(\ref{eq:lag_xi_full}) as
\be
{\cal L}_\xi\big|_{\rm 2nd} &=& 
- \frac{1}{2} A_a^{\rm df}\left( \delta_{ab}\p^2 + \delta_{ab}D^2 - D_b^\dagger D_a\right) A_b^{\rm df}
- \frac{1}{2} A_a^{\rm d}\left( \delta_{ab}\p^2 +\xi D_a D_b^\dagger\right) A_b^{\rm d}.
\label{eq:2ndline}
\ee
Now we see that the divergence-free part $A_a^{\rm df}$ 
does not contain the gauge-fixing parameter $\xi$, 
whereas the divergence part $A_a^{\rm d}$ depends on $\xi$, rendering it  an unphysical degree of freedom.

We can rewrite the divergence part of the Lagrangian 
to make the mass spectra of $A_a^{\rm d}$ explicit. 
Using Eq.~(\ref{eq:div_ex}) we obtain from the second term of 
Eq.~(\ref{eq:2ndline}) and the expansion (\ref{eq:exp_div_6}) 
without the $n=0$ part 
\be
\int d^{D-4}x\, {\cal L}_\xi\big|_{\rm 2nd}^{\rm d}
&=&-\frac{1}{2}\int d^{D-4}x\,
 \frac{1}{2} K D^{-2} \left(\p^2 + \xi D^2\right) K
\nonumber \\
&=& - \frac{1}{2} \sum_{n\neq0} K^{(n)} \left(\p^2 
+ \xi m_n^2 \right) K^{(n)}.
\label{eq:L_extra_div}
\ee
Absence of the massless mode ($n=0$) is physically important in a low energy effective theory, as we will discuss in Sec.~\ref{sec:5d}.

In contrast to the divergence part, the divergence-free 
part makes sense only for $D\ge6$, since it does not exist
in $D=5$. Let us rewrite the first part 
of Eq.~(\ref{eq:2ndline}) as
\be
{\cal L}_\xi\big|_{\rm 2nd}^{\rm df} = 
- 2 A_a^{\rm df}\left( \delta_{ab}\p^2 + H_{ab} \right)A_b^{\rm df},
\label{eq:2ndline_b}
\ee
where we have defined an operator
\be
H_{ab} = \delta_{ab}D^2-D_b^{\dagger}D_a.
\label{eq:mass_operator}
\ee
This operator is $N\times N$ matrix with the rank $N-1$, 
where we denote $N \equiv D-4$. 
For two extra-dimensions $N=2$, we have 
\begin{equation}
H={\cal D}^\dagger{\cal D}, 
\end{equation}
with ${\cal D} = (D_5, -D_4)$. Then we can define a ``superpartner" 
$\tilde H$ as 
\begin{equation}
\tilde H={\cal D}{\cal D}^\dagger=D_5D_5^\dagger +D_4 D_4^\dagger. 
\end{equation}
It is easy to verify that $H$ and $\tilde H$ have identical 
 spectra except possible zero modes and the construction 
can be generalized to higher dimensions, as described 
in App.~\ref{app:higher_dim}.

Let us parametrize the 
eigenvectors of $\tilde H$ as
\begin{equation}
\vec A^{\rm df}(x,y) = \frac{1}{2}\begin{pmatrix}D_5^{\dagger} \\ 
-D_4^{\dagger}\end{pmatrix} \left(\Omega^{-1}\bar\phi_0(y) 
\bar K^{(0)}(x) + \bar D^{-2}\bar K(x,y)\right)\,,
\label{eq:div_free_KKmode}
\end{equation}
where $\bar K(x,y)$ is orthogonal to zero modes $\{\bar\phi_0\}$  
of $\bar D^2$. 
In this way, Eq.~(\ref{eq:2ndline_b}) becomes
\be
{\cal L}_\xi\big|_{\rm 2nd}^{{\rm df},D=6} 
&=&  - \frac{1}{2\Omega^2} \bar\phi_0 \bar D^2\left(\p^2 + 
\bar D^2\right)\bar\phi_0 
- \frac{1}{2} \bar K \bar D^{-2} \left( \p^2 + \bar D^2 \right) 
\bar K \nonumber\\
&=& - \frac{1}{2} \bar K \bar D^{-2} \left( \p^2 + \bar D^2 \right) \bar K.
\label{eq:L_extra_div_free}
\ee
It is important to realize that the zero modes of $\bar D^2$ {\it always} disappear from the physical spectrum.
Now, it is natural to expand $\bar K$ in terms of the 
eigenfunctions of $\bar D^2$ as
\be
\bar K(x,y) = \sum_{n\neq0} \bar m_n \bar\phi_n(y) \bar K^{(n)}(x),
\ee
with the mass dimensions 
$\left[\bar K\right] = D/2$
and  
$\left[\bar K^{(n)}\right] = 1$.
Plugging this into Eq.~(\ref{eq:L_extra_div_free}) and integrating 
it over the extra dimensions, we get
\be
\int d^2x\, {\cal L}_\xi\big|_{\rm 2nd}^{{\rm df},D=6} =  
- \frac{1}{2} \sum_{n\neq0}  \bar K^{(n)}  
\left(\p^2 + \bar m_n^2 \right)  \bar K^{(n)}.
\label{eq:eff_Lag_df_6d}
\ee
This gives us the spectrum of the divergence-free part for two extra-dimensions.

Contrary to $D=6$, it is not easy to diagonalize $H$ for $D\ge 7$.
We will  study $D=7$ in Sec.~\ref{sec:monopole} for 
a spherically symmetric background.
We leave analysis of generic $D \ge 7$ as a future problem.

Absence of  zero modes in the extra-dimensional components, 
which we have explicitly shown  in $D=5$ and 6, sounds physically 
natural in the following sense. 
The non-trivial $\beta(y)$ implies that the gauge coupling depends 
on the extra-dimensional coordinate. 
This seemingly contradicts  $D$-dimensional gauge symmetry, 
and only the four-dimensional gauge symmetry holds. 
While the four-dimensional gauge symmetry ensures the existence 
of the massless four-dimensional gauge field as in 
Eq.~(\ref{eq:L1st_4dim}), there is no symmetric reasons explaining 
zero modes in the extra-dimensional components. 
This physical intuition makes the absence of zero modes plausible 
for $A_a$ in generic $D$, although a rigorous proof is lacking. 
We will partially verify this for $D=7$ in Sec.~\ref{sec:monopole}. 

Let us mention, however, that for \emph{separable} potentials, 
say $\beta = \beta_4(x^4) \beta_5(x^5)\cdots \beta_{D-1}(x^{D-1})$
corresponding to domain wall junctions,
we can understand the spectrum completely in a  recursive fashion, see
Appendix \ref{app:A} for details. 
We emphasize that there are no physical zero modes in the 
divergence-free components in the separable case in generic $D$ dimensions.

\subsection{Ghosts $c$ and $\bar c$}

Finally, we are  left with the third term of Eq.~(\ref{eq:lag_xi_full}) 
for the ghosts.
As before, we expand $c$ and $\bar c$ in terms of the 
eigenfunctions $\phi_n$ of the $D^2$ operator as
\be
c(x,y) = \sum_n \phi_n(y) c^{(n)}(x),
\ee
and similar for $\bar c$. 
Plugging these into the third term of Eq.~(\ref{eq:lag_xi_full}) and integrate it over the extra-dimensions, 
we get
\be
\int d^{D-4}\, {\cal L}_\xi\big|_{{\rm 3rd}} = - \sum_n \bar c^{(n)}\left(\p^2 + \xi m_n^2\right) c^{(n)}.
\ee

\subsection{Summary of KK decomposition}

Let us summarize the four-dimensional effective theory by 
gathering all the pieces obtained above.
The most relevant part for the low energy physics is massless 
fields. We found them in the four-dimensional gauge fields 
$A_\mu^{(0)}$
and the ghosts $c^{(0)}$ and $\bar{c}^{(0)}$. 
Their effective Lagrangian is given by
\be
{\cal L}_{\rm eff}^{(n=0)} = 
\frac{1}{2} A_\mu^{(0)} \left[
\eta^{\mu\nu}\p^2 - \left(1-\frac{1}{\xi}\right)\p^\mu\p^\nu 
\right] A_\nu^{(0)}- \bar c^{(0)} \p^2  c^{(0)}.
\ee
This is nothing but the ordinary four-dimensional Lagrangian 
for massless gauge fields  in the covariant gauge.
There is no other massless fields  
in our simple model given in Eq.~(\ref{eq:lag}), which we explicitly showed  for $D=5,6$ above, will show for $D=7$ in
Sec.\ref{sec:monopole}, and expect for $D\ge8$.
This is a virtue of our model which is in sharp contrast to other extra-dimensional models with compact
extra-dimensions where $A_a$ often supplies extra massless scalar fields
in the low energy physics. 
If one wants to avoid such massless scalars, one needs an 
additional assumption, for example, the $Z_2$ parity for $D=5$ 
model with $S^1/Z_2$ extra-dimension.

Next, we describe massive modes. We first collect the relevant pieces to describe 
the four-dimensional massive gauge fields,
$A_\mu^{(n)}$, the
divergence part of extra-dimensional gauge field $K^{(n)}$ defined in Eq.~(\ref{eq:exp_div_6}), and
the ghosts $c^{(n)}, \bar c^{(n)}$.
The effective Lagrangian takes the form
\be
{\cal L}_{{\rm eff};1}^{(n\neq0)} &=& 
\frac{1}{2} A_\mu^{(n)} \left[
\eta^{\mu\nu}\p^2 - \left(1-\frac{1}{\xi}\right)\p^\mu\p^\nu + \eta^{\mu\nu} m_n^2 \right] A_\nu^{(n)} \nonumber\\
&& -\, \frac{1}{2} K^{(n)}  \left(\p^2 + \xi m_n^2 \right)   K^{(n)}
- \bar c^{(n)}\left(\p^2 + \xi m_n^2\right) c^{(n)}.
\label{eq:eff_lag_n}
\ee
We find that the four-dimensional divergence part 
$\p^\nu A_\nu^{(n)}$, the extra-dimensional divergence part 
$K^{(n)}$, and the ghosts have the same mass square $\xi m_n^2$. 
The identical mass spectra assures that the contributions of 
divergence parts are cancelled by those of ghosts. 
We note a similarity of the Lagrangian (\ref{eq:eff_lag_n}) 
to that of four-dimensional $R_\xi$ gauge if we replace $K^{(n)}$ 
by the Nambu-Goldstone field \footnote{Similar result was recently reported for $D=5$ \cite{Okada:2017omx}.}. 
Our gauge-fixing condition (\ref{eq:gff}) is designed to exhibit 
this similarity explicitly. 
The physical degrees of freedom are the massive gauge field 
$A_\mu^{(n){\rm df}}$ with mass $m_n$.

For the divergence-free part of extra-dimensional components 
of gauge fields $A_a^{\rm df}$, we can explicitly write down 
effective Lagrangian for $D\le 6$. 
For $D=5$, the divergence free part does not exist. 
For $D=6$, we have one scalar field $\bar K$ whose KK modes 
defined in Eq.(\ref{eq:div_free_KKmode}) obey the effective Lagrangian 
(\ref{eq:eff_Lag_df_6d}). 
With the $\xi$-indenpendent masses, ${\bar K}^{(n)}$ ($n\neq0$) are 
physical degrees of freedom. 
For the higher dimensions $D\ge 7$, we anticipate $D-5$ KK 
towers of physical scalar fields. 
We can construct the full spectrum recursively for fully 
separable $\beta$, see App.~\ref{app:A}. 

\section{Examples}
\label{sec:3}

\subsection{Domain walls in $D=5$}
\label{sec:5d}

\subsubsection{A simple gauge kinetic function}
\label{sec:DW_simple}

In this subsection, we investigate localized modes of gauge 
fields on a domain wall in $D=5$. 
Although, generic results of the previous section are all valid 
in any dimension, it is worthwhile to illustrate the 
analysis in $D=5$ explicitly, as it is the simplest case.

We begin with a classical Lagrangian
\be
{\cal L} &=& - a^2 \varphi^2 {\cal F}_{MN}{\cal F}^{MN} + 
{\cal L}_{\rm kink},\\ 
{\cal L}_{\rm DW} &=& \p_M\sigma \p^M\sigma + \p_M\varphi\p^M\varphi 
- \Omega^2 \varphi^2
- \lambda^2 \left(\sigma^2 + \varphi^2 - v^2\right)^2.
\label{eq:L_DW}
\ee
Here, $\varphi$ and $\sigma$ are real scalar fields. 
The scalar field $\sigma$ is responsible for having a domain 
wall while $\varphi$ localizes gauge fields on the domain wall.
There are two discrete vacua $(\sigma,\varphi) = (\pm v,0)$. 
If we assume $\lambda v > \Omega$ the domain wall interpolating 
between them reads:
\be
\sigma= v\tanh\Omega y,\quad
\varphi =   \pm \bar v\,\sech\,\Omega y,\quad
\bar v \equiv \sqrt{v^2 - \frac{\Omega^2}{\lambda^2}}.
\label{eq:sol_DW}
\ee

With this solution as the background configuration, the relevant 
part of Lagrangian for the gauge field is given by
\be
{\cal L}_A = - \beta(y)^2 {\cal F}_{MN} {\cal F}^{MN},
\quad \beta(y) = a\bar v\,\sech\,\Omega y.
\ee
The differential operators associated this background solution are 
\be
D^2 &=& D_y^\dagger D_y = - \p_y^2 + \Omega^2 \left(1 - 2\, 
\sech^2\,\Omega y\right),\label{eq:D2_kink}\\
\bar D^2 &=& D_y D_y^\dagger = - \p_y^2 + \Omega^2.
\ee
Operators $D^2$ and $\bar D^2$ can be regarded as components of the 
Hamiltonians of supersymmetric quantum mechanics in one dimension.
Therefore, the energy eigenvalues are identical except for 
zero eigenvalue. 
Since $\bar D^2$ has no  zero modes, the general solution 
corresponding to the zero eigenvalue 
\be
\bar\phi_0 = A e^{\Omega y} + B e^{-\Omega y},\quad m_0 = 0
\ee 
is not normalizable.
The eigenfunctions of physical states in the continuum are 
\be
\bar\phi(y;k) = \frac{e^{iky}}{\sqrt{2\pi\Omega}} ,\quad m(k) 
= \sqrt{k^2 + \Omega^2}\,,
\ee
where the normalization is chosen as
\be
\int^\infty_{-\infty} dy\, \bar\phi(y;k)^* \bar\phi(y;k') 
= \frac{1}{\Omega} \delta(k-k').
\ee
On the other hand, general solution for zero eigenvalue of $D^2$ 
is given by
\be
\phi_0(y) = \sqrt{\frac{\Omega}{2}}\, \sech\,\Omega y 
+ B \left(\sinh \Omega y+y \Omega\,  \sech\, \Omega y\right),\quad m_0 = 0.
\ee
Since the second term diverges at $y=\pm \infty$, we take $B=0$. 
\begin{figure}[t]
\begin{center}
\includegraphics[height=6cm]{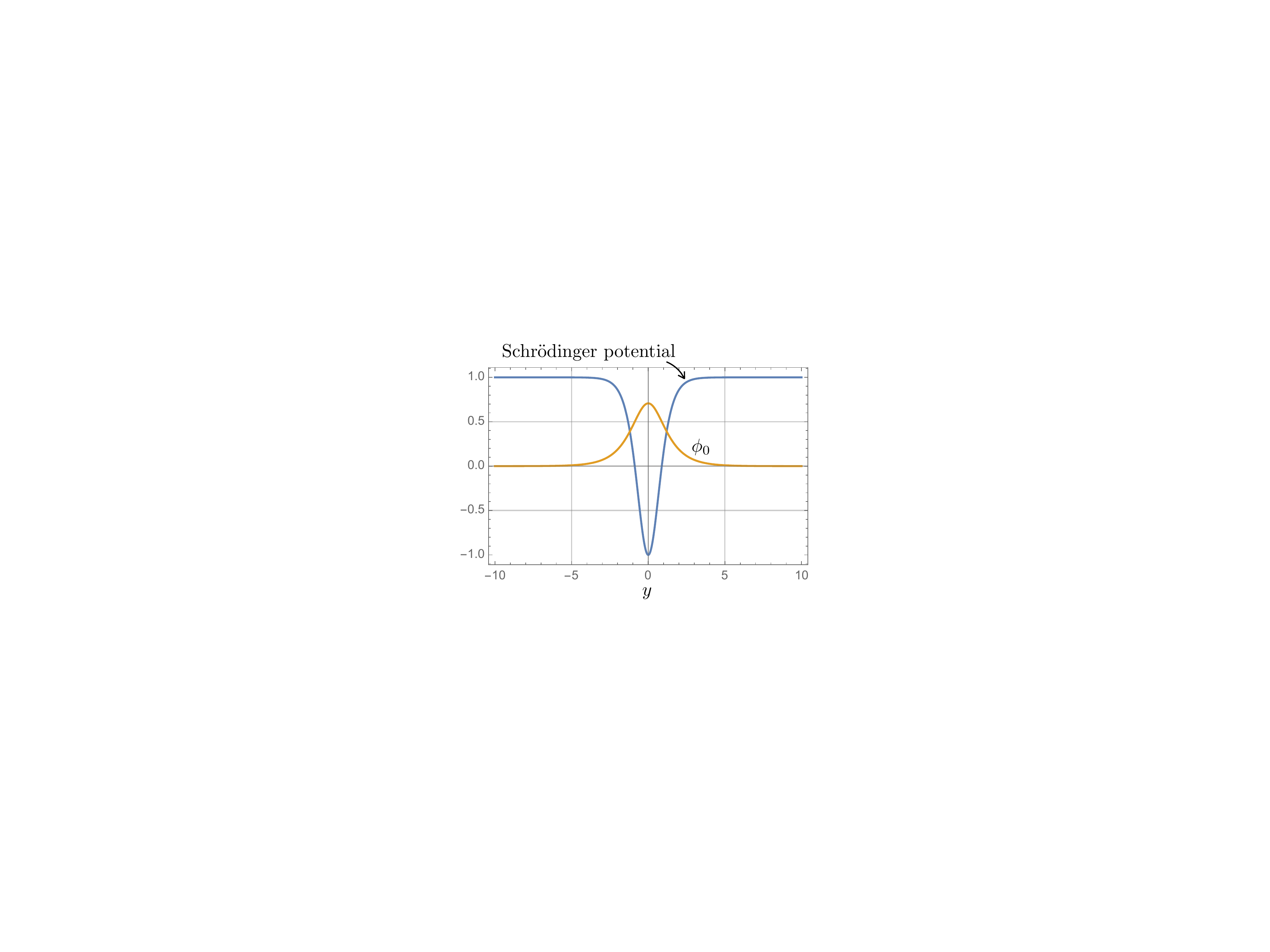}
\caption{The Schr\"odinger potential and its zero mode $\phi_0$.}
\label{fig:schV_kink}
\end{center}
\end{figure}
Fig.~\ref{fig:schV_kink} shows the Schr\"odinger potential given 
in Eq.~(\ref{eq:D2_kink}) and the zero mode $\phi_0$.
The physical continuum of $D^2$ eigenstates are 
obtained by supersymmetry between $D^2$ and $\bar{D}^2$ which relates
\be
D_y^\dagger\bar{\phi}(y;k)=-m(k)\phi(y;k), \quad 
D_y\phi(y;k)=-m(k)\bar{\phi}(y;k).
\label{SUSY1}
\ee
From this, we have
\be
\phi(y;k) = \frac{-1}{m(k)} D_y^\dagger \bar \phi(y;k) = \frac{-i}{m(k)} e^{iky}\left(k + i \Omega \tanh \Omega y\right),
\quad
m(k) = \sqrt{k^2 + \Omega^2}\,.
\ee
The threshold is 
$\phi(y;k=0) = \tanh \Omega y$.
The normalization reads
\be
\int^\infty_{-\infty} dy\, \phi(y;k)^* \phi(y;k') 
= \int^\infty_{-\infty} dy\, \frac{1}{m(k)^2} \bar\phi(y;k)^* \bar D^2 \bar \phi(y;k') 
= \frac{1}{\Omega} \delta(k-k').
\ee

The quantum Lagrangian given in Eq.~(\ref{eq:lag_xi_full}) 
takes the form
\be
{\cal L}_\xi &=& \frac{1}{2} A_\mu \left[
\eta^{\mu\nu}\p^2 - \left(1-\frac{1}{\xi}\right)\p^\mu\p^\nu + \eta^{\mu\nu}D^2\right] A_\nu \nonumber\\
&&-\,\frac{1}{2} A_y\left(
\p^2 - \xi \bar D^2 \right) A_y
- \bar c \left(\p^2 + \xi D^2\right)c.
\label{eq:L5}
\ee
We expand $A_\nu$  and $c$ in terms of $\phi_0(y)$ and $\phi(y;k)$ as
\be
A_\nu(x,y) &=& \phi_0(y) A_\nu^{(0)}(x) + \int_{-\infty}^\infty dk\, A_\nu(x;k) \phi(y,k) ,\\
c(x,y) &=& \phi_0(y) c^{(0)}(x) + \int_{-\infty}^\infty dk\, c(x;k) \phi(y,k),
\ee
and similarly for $\bar c$. Here, $A_\nu(x;k)^* = A_\nu(x;-k)$ is imposed.
On the other hand, we expand $A_y$ in terms of eigenfunctions of $\bar D^2$ operator:
\be
A_y(x,y) = \int_{-\infty}^\infty dk\, A_y(x;k) \bar \phi(y,k) ,
\label{eq:Ay}
\ee
with $A_y(x;k)^* = A_y(x;-k)$.
Therefore, the absence of massless modes in $A_y$ is a direct 
consequence of absence of physical zero modes 
in $\bar D^2$. 
For illustration, let us compare the simple expansion here and 
one based on the generic arguments around Eq.~(\ref{eq:div_ex}). 
We first write
$A_y = D_y D^{-2} K$
and expand
$K$ in terms of $\phi(y;k)$.
The basis for the expansions are different from Eq.~(\ref{eq:Ay}), nevertheless, we get the same four-dimensional 
Lagrangian by using (\ref{SUSY1}).
We can express divergence of Eq.~(\ref{eq:Ay}) as
\be
D_y^\dagger A_y(x,y) = -\int_{-\infty}^\infty dk\, m(k) A_y(x;k) \phi(y;k).
\ee
This is nothing but the counterpart of (\ref{eq:exp_div_6}).

Plugging the expansions above into Eq.~(\ref{eq:L5}) and integrating it over $y$, we get
\be
{\cal L}_\xi^{\rm eff} = {\cal L}_\xi^{(0)} + \int_{-\infty}^\infty \frac{dk}{\Omega}\, {\cal L}_\xi(k),
\ee
where we have the massless part
\be
{\cal L}_\xi^{(0)} 
= \frac{1}{2}A_\mu^{(0)}\left[\eta^{\mu\nu}\p^2 - \left(1-\frac{1}{\xi}\right)\p^\mu\p^\nu \right] A_\nu^{(0)} 
- \bar c^{(0)}\p^2 c^{(0)},
\label{eq:Lxi_0}
\ee
and  the massive parts
\be
{\cal L}_\xi(k) 
&=& \frac{1}{2} A_\mu(x;k)^* \left[\eta^{\mu\nu}\p^2 
- \left(1-\frac{1}{\xi}\right)\p^\mu\p^\nu + \eta^{\mu\nu}m(k)^2 \right]  
A_\nu(x;k)  \nonumber\\
&&-\,   \frac{1}{2} A_y(x;k)^*  \left(\p^2 + \xi m(k)^2 \right) A_y(x;k) 
- {\bar c}(x;k) \left(\p^2 + \xi m(k)^2\right) c(x;k).
\label{eq:Lxi_n}
\ee
Thus, we conclude that the low energy effective theory on the domain wall in $D=5$ includes
one massless gauge field $A_\mu^{(0)}(x)$ and the continuum KK towers of massive vector fields with
the mass gap $\Omega$. We emphasise that the absence of other massless modes is not an assumption
but a logical consequence.

It would be useful to rewrite the above effective Lagrangians into the standard form.
For the massless fields, our model in Eq.~(\ref{eq:Lxi_0}) can be expressed as  
\be
{\cal L}_\xi^{(0)} = - \frac{1}{4} F_{\mu\nu}^{(0)} F^{(0)\mu\nu} 
- \frac{1}{2\xi} f^{(0)}{}^2 - \bar c^{(0)}\p^2 {c}^{(0)},
\label{eq:L_eff_0}
\ee
with
\be
 f^{(0)} = \p^\mu A_\mu^{(0)}.
\ee
Similarly,  (\ref{eq:Lxi_n}) can be expressed as
\be
{\cal L}_\xi(k)
&=& - \frac{1}{4}F_{\mu\nu}(k)^* F^{\mu\nu}(k) 
+\, \frac{1}{2} | \p_\mu  A_y(k) - m(k) A_\mu(k)
|^2 \nonumber\\
&&-\, \frac{1}{2\xi} |f(k)|^2
- \bar c(k) \left(\p^2 + \xi m(k)^2\right) c(k),
\label{eq:L_eff_n}
\ee
where we abbreviated $A_\mu(k) \equiv A_\mu(x;k)$. 
The gauge fixing function is given by
\be
f(k) = \p^\mu A_\mu(k) + \xi m(k)  A_y(k).
\label{eq:gf_functionals_eff}
\ee
It is now quite clear that
$A_y(k)$ is a St\"uckelberg-like field which pretends to be  
a Nambu-Goldstone field absorbed by the gauge field via 
the Higgs mechanism.
The effective Lagrangians in  (\ref{eq:L_eff_0}) and (\ref{eq:L_eff_n}) 
are the result of dynamical compactification of the infinitely 
large fifth dimension by the domain wall. 

\begin{figure}[t]
\begin{center}
\includegraphics[width=12cm]{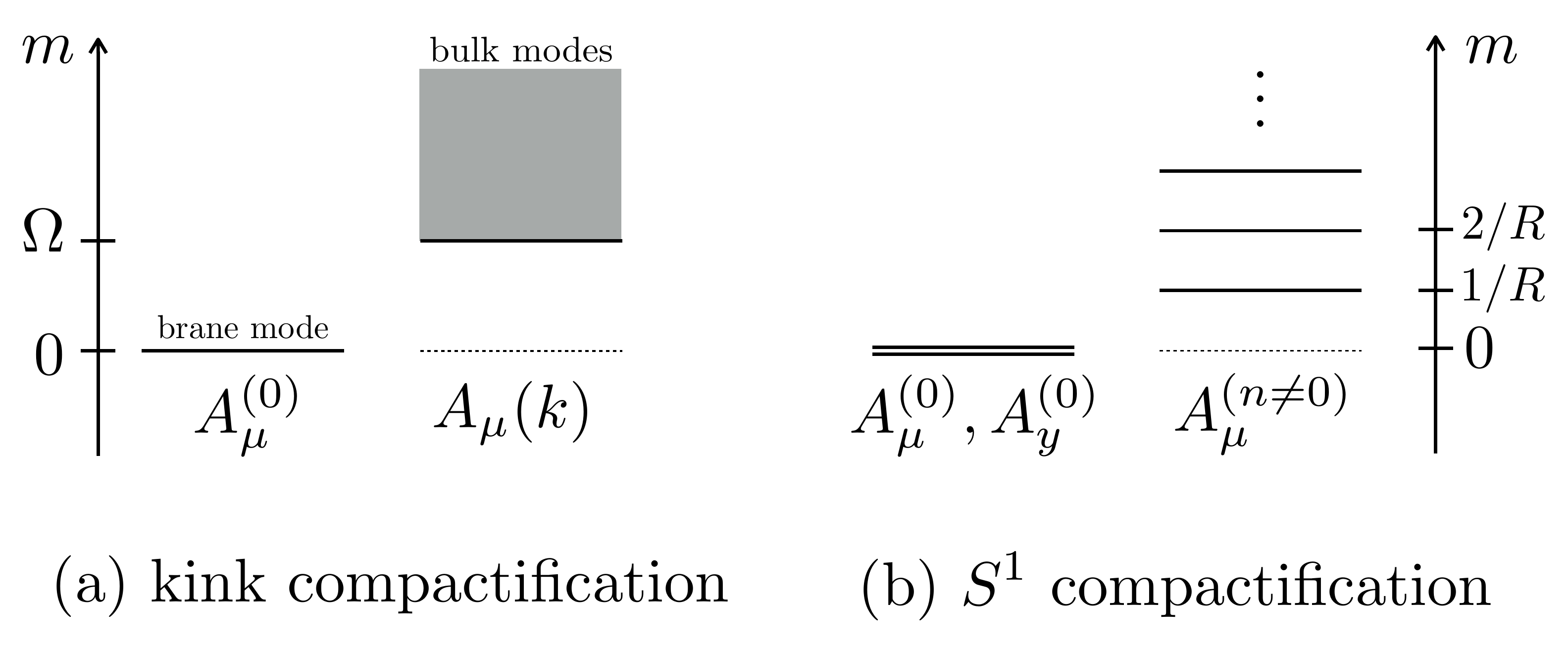}
\caption{KK mass spectra of physical fields for (a) our model and (b) $S^1$ extra-dimension.}
\label{fig:spctra}
\end{center}
\end{figure}

Let us now compare this with the model, where the extra-dimension 
is compactified by hand to a circle $S^1$ of the radius $R$. 
The most important difference is in the massless fields. Domain-wall compactification produces only massless four-dimensional gauge fields, whereas
the $S^1$ model has in addition a massless scalar field originating from $A_y$. 
One cannot avoid this scalar because all modes are normalizable when the extra-dimension is compact. 
In order to suppress it, one need additional instruments such as $Z_2$ orbifolding and
parity conditions, and so on. 
In contrast, the massive modes of both models are quite similar. 
In both models there exists a mass gap,
the inverse width $\Omega$ of the domain wall and the inverse 
radius $1/R$, respectively. 
Above the mass gap the domain wall model has a continuum spectrum, 
while in $S^1$ model there is an equidistant discrete tower 
of massive modes, which is the compact version of the continuum.
A more important difference is that in our model the massless gauge fields are localized on the domain wall,
and all the other massive fields are bulk fields. On the other hand, all fields (both massless and massive) are
spread uniformly across the entire extra-dimension in the $S^1$ model. Fig.~\ref{fig:spctra} summaries the differences.

Before closing this section, let us make a comment on the gauge transformation.
Note that the gauge transformation for the original gauge field, ${\cal A}_M \to {\cal A}_M' =  {\cal A}_M + \p_M \Lambda$ is 
translated for the canonically normalized fields as $A_M' = A_M + 2\beta \p_M \Lambda$.
Expanding the gauge transformation parameter as
\be
\Lambda = 
\frac{\phi_0(y)}{2\beta}  \Lambda^{(0)}(x) + 
\int^\infty_{-\infty}dk\, \frac{1}{2\beta} \phi(y;k)\Lambda(x;k),
\ee
with $\Lambda(x;k)^* = \Lambda(x;-k)$,
we find
\be
A_\mu^{(0)} &\to& A_\mu^{(0)}{}' = A_\mu^{(0)} + \p_\mu \Lambda^{(0)},\\
A_\mu(k) &\to& A_\mu(k)' = A_\mu(k) + \p_\mu \Lambda(k),\\
A_y^{(0)} &\to& A_y^{(0)}{}' = A_y^{(0)},\label{eq:gtf_Ay}\\
 A_y(k) &\to&  A_y(k)' =  A_y(k) + m(k) \Lambda(k).\label{eq:gtf_Ay_n}
\ee
For example, the gauge transformation of $A_y(x,y)$ can be obtained as follows
\be
A_y' &=& \bar\phi_0 A_y^{(0)}
+ \int_{-\infty}^\infty dk\,  A_y(x;k) \bar \phi(y;k) \nonumber\\
&&+\, 2\beta \p_y \left[\frac{\phi_0(y)}{2\beta}  \Lambda^{(0)}(x) + 
\int^\infty_{-\infty}dk\, \frac{1}{2\beta} \phi(y;k)\Lambda(x;k) \right]\nonumber\\
&=& \bar\phi_0 A_y^{(0)} 
- \int^\infty_{-\infty}dk\, \frac{1}{m(k)}D_y \phi(y;k) \left(A_y(k) + m(k) \Lambda(x;k)\right),
\ee
where we have kept $\bar\phi_0$, although it is unphysical, 
and we have used the fact $\phi_0 = \beta$, $D_y = - \beta \p_y\beta^{-1}$ 
and 
(\ref{SUSY1}).

Note that the gauge transformation law (\ref{eq:gtf_Ay_n}) correctly 
derives the ghost Lagrangian in Eq.~(\ref{eq:L_eff_n})  as 
variations of the gauge fixing function given in 
Eq.~(\ref{eq:gf_functionals_eff}).

\paragraph*{Remark:}

All the results obtained in this subsection are consistent with 
our previous works on domain walls in flat 5 dimensions 
\cite{Ohta:2010fu,Arai:2012cx,Arai:2013mwa,Arai:2014hda,Arai:2016jij,Arai:2017lfv,Arai:2017ntb}. 
In particular, the absence of $A_y^{(0)}$ is one of the 
important physical results. 
However, the previous analysis in \cite{Ohta:2010fu,Arai:2012cx,Arai:2013mwa,Arai:2014hda,Arai:2016jij,Arai:2017lfv,Arai:2017ntb} 
were carried out in the axial gauge $A_y = 0$. 
Although the axial gauge is useful at least in classical analysis,
$A_y^{(0)}$ is not transformed by any gauge transformation as 
explicitly shown in Eq.~(\ref{eq:gtf_Ay}).
Therefore, the axial gauge $A_y = 0$ cannot exclude $A_y^{(0)}$. Therefore,
the analysis in this work justifies the absence of the  
zero mode $A_y^{(0)}$ in our previous works.

\subsubsection{More general gauge kinetic functions}

In the previous section, 
the only localized field is  the massless four-dimensional gauge fields.
All the massive modes are continuum bulk modes. When we want several massive bound states,
it can be realized, for example, as follows. We do not change the domain wall Lagrangian ${\cal L}_{\rm DW}$
in Eq.~(\ref{eq:L_DW}). 
Instead, we modify $\beta$ as
\be
{\cal L}_A = - (\beta^{(n)})^2 F_{MN}F^{MN},\quad
\beta^{(n)}(\varphi) \equiv a \varphi^n.
\ee
All the formulae given in Sec.~\ref{sec:DW_simple} remain the same if we replace $\beta$ by $\beta^{(n)}$.
The Hamiltonians are given by
\be
(D^{(n)})^2 &=& - \p_y^2 + V^{(n)},\quad V^{(n)} = n \Omega^2 \left(-1+(n+1)\tanh^2\Omega y\right),\\
(\bar D^{(n)})^2 &=& - \p_y^2 + \bar V^{(n)},\quad \bar V^{(n)} = n\Omega^2\left(1+(n-1)\tanh^2\Omega y\right).
\ee
The following descent relation holds
\be
\bar V^{(n)} = V^{(n-1)} + (2n-1)\Omega^2.
\label{eq:descent}
\ee
For any $n$, the physical modes reside
only in the four-dimensional part $A_\mu$. 
As can be seen from Eq.~(\ref{eq:L5}), the physical mass spectrum 
is determined by the $(D^{(n)})^2$ operator.
The zero mode is immediately found as $\beta^{(n)}$. 
On the other hand, the descent relation (\ref{eq:descent}) 
implies that 
the eigenfunction of $\bar V^{(n)}$ is in one-to-one 
correspondence with that of $V^{(n-1)}$, whose eigenvalue is 
shifted by $(2n-1)\Omega^2$. 
We can find excited modes of $(D^{(n)})^2$ from eigenfunctions of 
the superpartners $(\bar D^{(n)})^2$, which share the same non-zero 
eigenvalues, by applying $D^{(n)\dagger}$. 
Thus we can recursively construct all the discrete modes of 
$(D^{(n)})^2$ starting from the zero mode. 
For example, the first excited mode of $\bar V^{(n)}$ is 
$\beta^{(n-1)}$ which is the zero mode of $V^{(n-1)}$ and 
the mass squared is $(2n-1)\Omega^2$. 
The first excited state of $V^{(n)}$ can be obtained 
by multiplying $D^{(n)\dagger}$ 
on $\beta^{(n-1)}$. 

\begin{figure}[ht]
\begin{center}
\includegraphics[width=16cm]{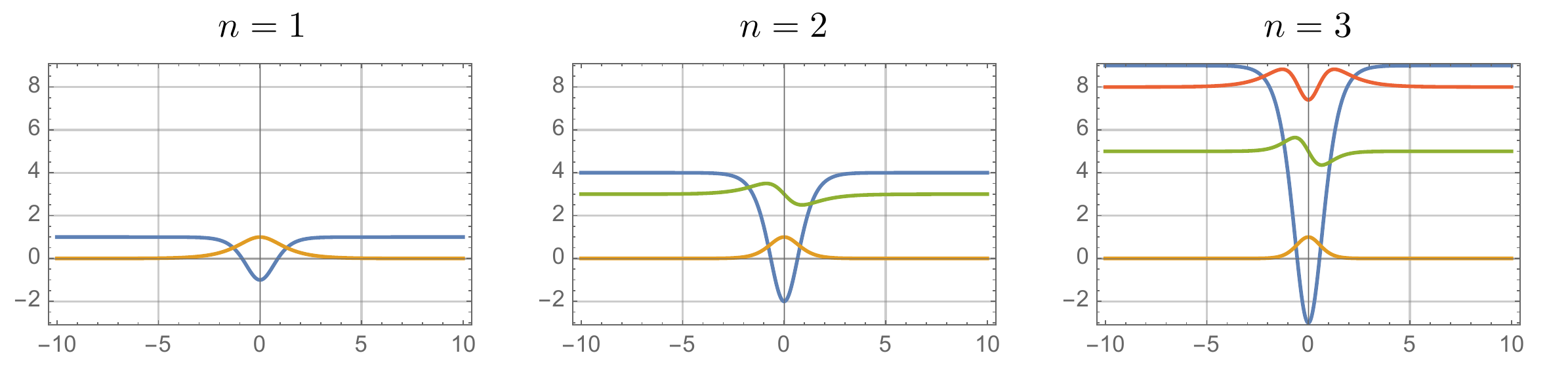}
\caption{The Schr\"odinger potentials $V^{(n)}$ and their bound states 
are shown for $n=1,2,3$. The horizontal axis is $\Omega y$ and
the vertical axis is $V^{(n)}/\Omega^2$.}
\label{fig:bound_KK}
\end{center}
\end{figure}

To illustrate the recursive procedure, let us 
consider $n=2$ with 
$\beta^{(2)} = a\, (\bar v\,\sech\,\Omega y)^2$.
We find two bound states
\be
\phi_0^{(2)} &\propto& \beta^{(2)} \propto \sech^2\Omega y,\quad m_0^2 = 0,\\
\phi_1^{(2)} &\propto& D^{(2)\dagger} \phi^{(1)}_0 
\propto \sech^2\Omega y \sinh \Omega y, \quad m_1^2 = 3\Omega^2,
\ee
with $\phi^{(1)}_0 \propto  \beta^{(1)}$.
The continuum bulk modes follow and their mass squares are given by
$m(k)^2 = k^2 + \Omega^2 + 3\Omega^2 = k^2 + 4\Omega^2$.
Similarly  
we can understand $n=3$:
\be
\phi_0^{(3)} &\propto& \beta^{(3)} \propto \sech^3\Omega y,\quad m_0^2 = 0,\\
\phi_1^{(3)} &\propto& D^{(3)\dagger} \phi^{(2)}_0 \propto 
\sech^3\Omega y \sinh \Omega y, \quad m_1^2 = 5\Omega^2,\\
\phi_2^{(3)} &\propto& D^{(3)\dagger} \phi^{(2)}_1 \propto  
\sech^3 \Omega y \left(-1 + 4\sinh^2\Omega y\right) , 
\quad m_2^2 = 8\Omega^2.
\ee
The threshold mass squared for the continuum modes is 
$\Omega^2 + 3\Omega^2 + 5\Omega^2 = 9 \Omega^2$.
The Fig.~\ref{fig:bound_KK} shows the bound states for $n=1,2,3$ cases.
The spectrum for generic $n$ is straightforwardly obtained.
Having the analytic solutions for the bound KK modes is 
useful for model building\footnote{We thank to Nobuchika Okada for this point. See also recent paper \cite{Okada:2017omx,Okada:2018von}.}.

\subsection{Vortex in $D=6$}
\label{sec:6d}

Compared to the wealth of models in five non-compact dimensions with domain walls, there has been a very few six-dimensional models, 
where localization of massless gauge fields on a topological soliton is realized.
To the best of our knowledge, we believe that the model presented in this section is 
the first successful model in flat spacetime (without gravity), 
where this is achieved.

\subsubsection{ANO vortex string in $D=6$}

According to the general strategy of Sec.~\ref{sec:2}, the field 
dependent kinetic term (\ref{eq:lag}) with nontrivial $\beta$ 
in \emph{any} dimensions generates massless gauge fields in 
four-dimensional low-energy effective theory on a topological 
soliton. Here, we will give a concrete model in flat six dimensions, 
where an Abrikosov-Nielsen-Olsen (ANO) vortex is used to localize 
massless gauge fields.

We consider the following $U(1)\times \tilde U(1)$ model
\be
{\cal L} &=& - a^2 \varphi^2 {\cal F}_{MN}{\cal F}^{MN} + 
{\cal L}_{\rm vortex},\\ 
{\cal L}_{\rm vortex} &=& -\frac{1}{4\tilde e^2} \tilde F_{MN} 
\tilde F^{MN} + |D_M\sigma|^2 + (\p_M\varphi)^2 - V,\\
V &=& \frac{\lambda_1}{4}(|\sigma|^2 - v_1^2)^2 
+ \frac{\lambda_2}{4}(\varphi^2 - v_2^2)^2 + 
\lambda_3(|\sigma|^2 - v_1^2)(\varphi^2 - v_2^2).
\ee
The field strengths are given as ${\cal F}_{MN} = \p_M {\cal A}_N 
- \p_N {\cal A}_M$ and $\tilde F_{MN} = \p_M\tilde A_N - \p_N \tilde A_M$. 
The complex scalar field $\sigma$ is charged under $\tilde U(1)$, 
i.e. $D_M \sigma = (\p_M + i \tilde A_M) \sigma$, while $\varphi$ 
is a real scalar field. 

When $\sigma$ develops a non zero expectation value, $\tilde U(1)$ 
gauge symmetry is broken.
As a consequence, an ANO-type vortex is formed. The parameters 
of the potential $V$ are chosen such that $\varphi$ condenses 
only inside the ANO vortex, which leads to the localization of 
massless gauge fields. This is similar to 
superconducting cosmic strings \cite{Witten:1984eb,Frere:2000dc}, where 
$\varphi$ is complex and is charged under $U(1)$, so that the 
ANO string becomes superconducting. In our model, the vortex 
is not superconducting because $\varphi$ is neutral. 
Instead, it couples to the $U(1)$ gauge field ${\cal A}_M$ via 
the nontrivial field dependent gauge kinetic term. 

\begin{figure}[t]
\begin{center}
\includegraphics[height=6.5cm]{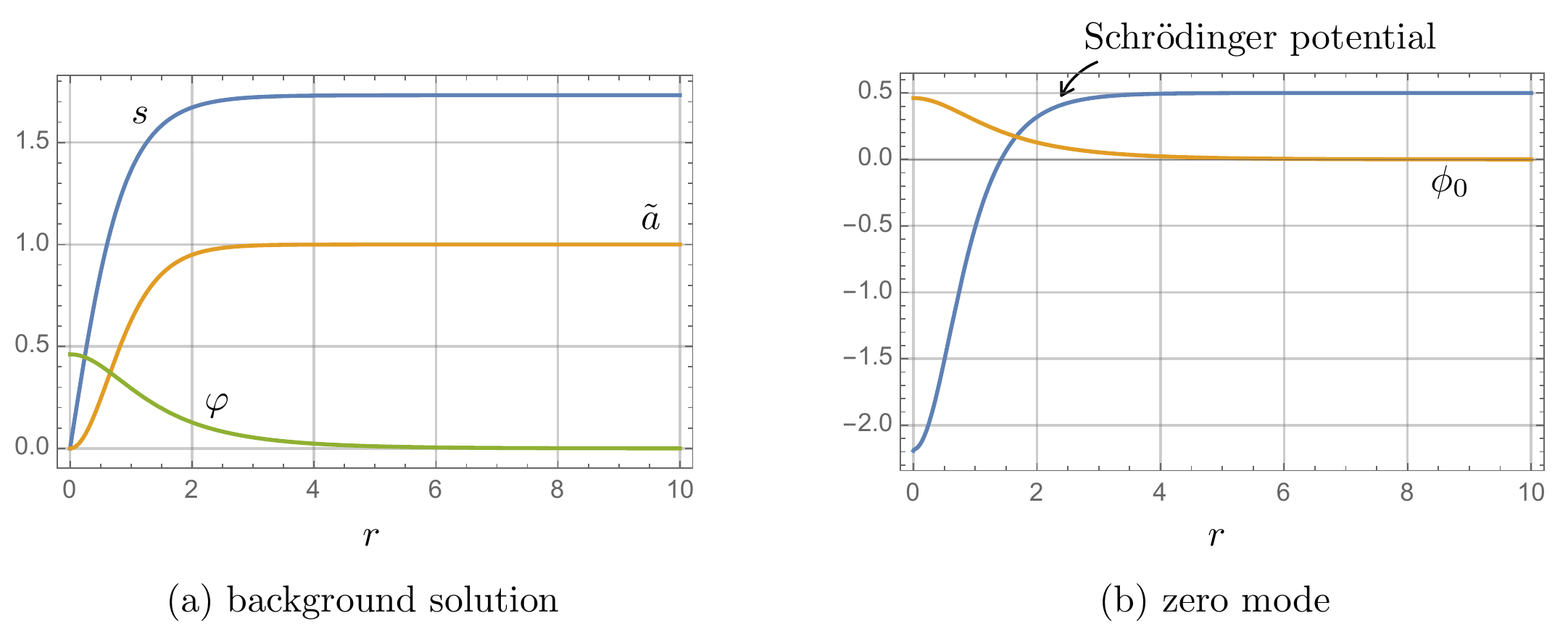}
\caption{The panel (a) shows the profile functions for a typical 
numerical solution of the single ANO vortex. The panel (b) shows 
the zero mode wave function of the gauge fields and the trapping 
Schr\"odinger type potential.}
\label{fig:schV_vortex}
\end{center}
\end{figure}

To find the background vortex solution let us make an Ansatz 
\be
\sigma = s(r) e^{i\theta},\quad \varphi = \varphi(r),
\quad \tilde A_a = \epsilon_{ab} \frac{x^b}{r^2}\tilde a(r),
\ee
where $r = \sqrt{x_4^2 + x_5^2}$ and $\theta = \arctan x_5/x_4$. 
A suitable boundary condition for the vortex is 
$s = \varphi'=\tilde a = 0$ at $r=0$ and $s = \sqrt{v_1^2 + 
\frac{2\lambda_3}{\lambda_1}v_2^2}$, $\varphi=0$ and $\tilde a = 1$ 
at $r = \infty$. 
A typical solution (obtained by numerical integration of equations 
of motion) is shown in Fig.~\ref{fig:schV_vortex} for 
$(\lambda_1,\lambda_2,\lambda_3)=(1,3,1)/v_1$, $v_2=v_1$, 
$\tilde e = v_1^{-1/2}$. 
As desired, the real scalar field $\varphi$ condenses around 
the vortex. Therefore $\beta^2 = a^2 \varphi^2$ is square 
integrable, which ensures localization 
of the massless $U(1)$ gauge field.

\subsubsection{The physical spectrum}

Let us next study the KK spectrum for the vortex background 
obtained above. 
Following the generic arguments in Sec.~\ref{sec:2}, the 
relevant equations can be read from Eq.~(\ref{eq:lag_xi_full}) as 
\be
&&
\left[\eta^{\mu\nu}\p^2 - \left(1-\frac{1}{\xi}\right)\p^\mu\p^\nu 
+ \eta^{\mu\nu}D^2\right] A_\nu = 0,\\
&&
\left[
\delta_{ab}(\p^2 + D^2) - \left(D_b^\dagger D_a 
- \xi D_a D^\dagger_b\right)\right] A_b = 0,\\
&&\left(\p^2 + \xi D^2\right)c = 0.
\ee
In what follows, we will concentrate on the physical modes: the 
transverse modes of $A_\mu$ and the divergence-free 
part of $A_a$. The transverse condition $\p^\mu A_\mu^{\rm T} = 0$ 
and the divergence-free condition $P_{ab}A_b^{\rm df} = 0$ give 
us the following equations 
\be
(\p^2 + D^2)A_\mu^{\rm T} = 0,\\
\left[(\p^2 + D^2)\delta_{ab} - D_b^\dagger D_a\right] A_b^{\rm df} = 0.
\label{eq:df_d=6}
\ee
Since the divergence-free part in $D=6$ can be expressed 
by means of $\bar K$ (see Eq.~(\ref{eq:div_free_KKmode})) as 
$
A_a^{\rm df} = \frac{1}{2}\epsilon_{ab}D^\dagger_b \bar D^{-2} \bar K
$
we find 
\be
\left(D^2\delta_{ab} - D_b^\dagger D_a\right) A_b^{\rm df} 
= \frac{1}{2}\epsilon_{ab}D^\dagger_b \bar K,
\ee
where we have omitted the zero mode for $A_a^{\rm df}$ since it does not appear in a physical spectrum.
Plugging this into Eq.~(\ref{eq:df_d=6}), we find the following 
equation for the divergence-free part
\be
\epsilon_{ab}D^\dagger_b \bar D^{-2}\left(\p^2 + \bar D^2\right) \bar K = 0.
\ee
In short, we just need to find the eigenvalues of the operator 
$D^2$ for $A_\mu^{\rm T}$ and $\bar D^2$ for $A_a^{\rm df}$.

\begin{figure}[ht]
\begin{center}
\includegraphics[width=16cm]{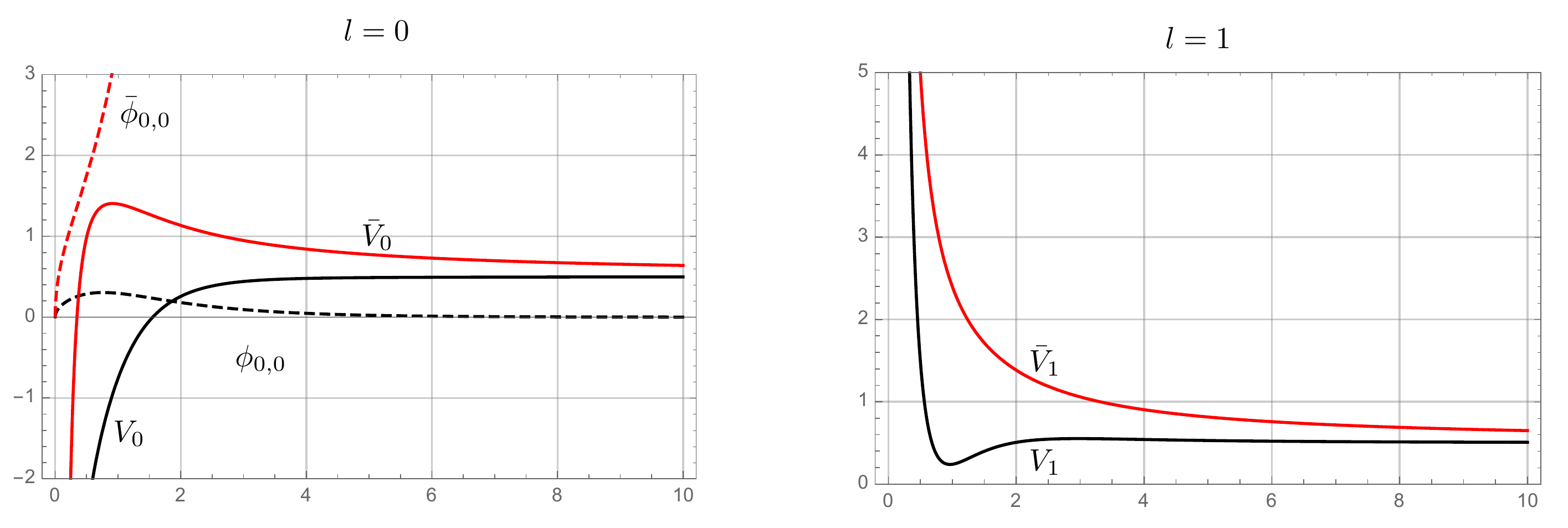}
\caption{The left panel shows $V_0$ (black solid curve) and 
$\bar V_0$ (red solid curve). The broken lines correspond to 
the zero modes $\phi_{0,0}$ (black) and $\bar \phi_{0,0}$ (red), 
respectively. 
Note that $\phi_{0,0}$ is related to $\phi_0$ in 
Fig.~\ref{fig:schV_vortex} by $\phi_{0,0} = \sqrt{r}\phi_0$.
The right panel shows $V_1$ and $\bar V_1$.}
\label{fig:eff_pot_vor_01}
\end{center}
\end{figure}

Let us next consider axially symmetric background with $\beta = \beta(r)$.
We expand a function of $x^\mu$, $x^4 = r \cos\theta$ and 
$x^5 = r\sin\theta$ as 
\be
f(x,r,\theta) = \sum_{n,l} f_{n,l}(x) \frac{\phi_{n,l}(r)
e^{i l\theta}}{\sqrt r},\quad l \in \mathbb{Z}.
\ee
Then we find the eigenvalue equations
\be
D^2\ &:&\ \left(-\frac{d^2}{dr^2} + V_l\right) \phi_{n,l} 
= m_{n,l}^2 \phi_{n,l},\\
\bar D^2\ &:&\ \left(-\frac{d^2}{dr^2} + \bar V_l\right) \bar \phi_{n,l} 
= \bar m_{n,l}^2 \bar\phi_{n,l},
\ee
with
\be
V_l &=& \frac{1}{\sqrt{r} \beta}\left(\sqrt{r} \beta\right)'' 
+ \frac{l^2}{r^2},\\
\bar V_l &=& \frac{1}{\sqrt{r}\beta^{-1}}\left(\sqrt{r} 
\beta^{-1}\right)'' + \frac{l^2}{r^2}.
\ee
The zero modes of both operators are $\phi_{0,0} = \sqrt{r} \beta$ 
and $\bar\phi_{0,0} = \sqrt{r}\beta^{-1}$.
Note that we can again rewrite the above equations in the SUSY 
QM fashion as
\be
\left(Q_r^\dagger Q_r + \frac{l^2}{r^2}\right)\phi_{n,l} 
= m_{n,l}^2 \phi_{n,l},\\
\left(\bar Q_r^\dagger \bar Q_r + \frac{l^2}{r^2}\right)
\bar \phi_{n,l} = \bar m_{n,l}^2 \bar\phi_{n,l},
\ee
where we introduce
\be
Q_r = - \p_r + \left(\p_r\log \sqrt{r}\beta\right),\quad
Q^\dagger_r = \p_r + \left(\p_r\log \sqrt{r}\beta\right),\\
\bar Q_r = - \p_r + \left(\p_r\log \sqrt{r}\beta^{-1}\right),\quad
\bar Q^\dagger_r = \p_r + \left(\p_r\log \sqrt{r}\beta^{-1}\right).
\ee
The term $l^2/r^2$ is nothing but the centrifugal potential 
for the mode of angular momentum $l$. 
Fig.~\ref{fig:eff_pot_vor_01} shows $V_{l=0,1}$ and $\bar V_{l=0,1}$ 
for the numerical vortex solution given in Fig.~\ref{fig:schV_vortex}.
For modes with $l > 0$, the centrifugal force significantly 
lifts the potential near origin. Therefore, bound states, 
if exist, are pushed away from the origin.  
We will work out analytic solutions for a typical 
gauge kinetic function $\beta$ in Sec.~\ref{sec:analytic_D=6}.

\begin{figure}[ht]
\begin{center}
\includegraphics[width=16cm]{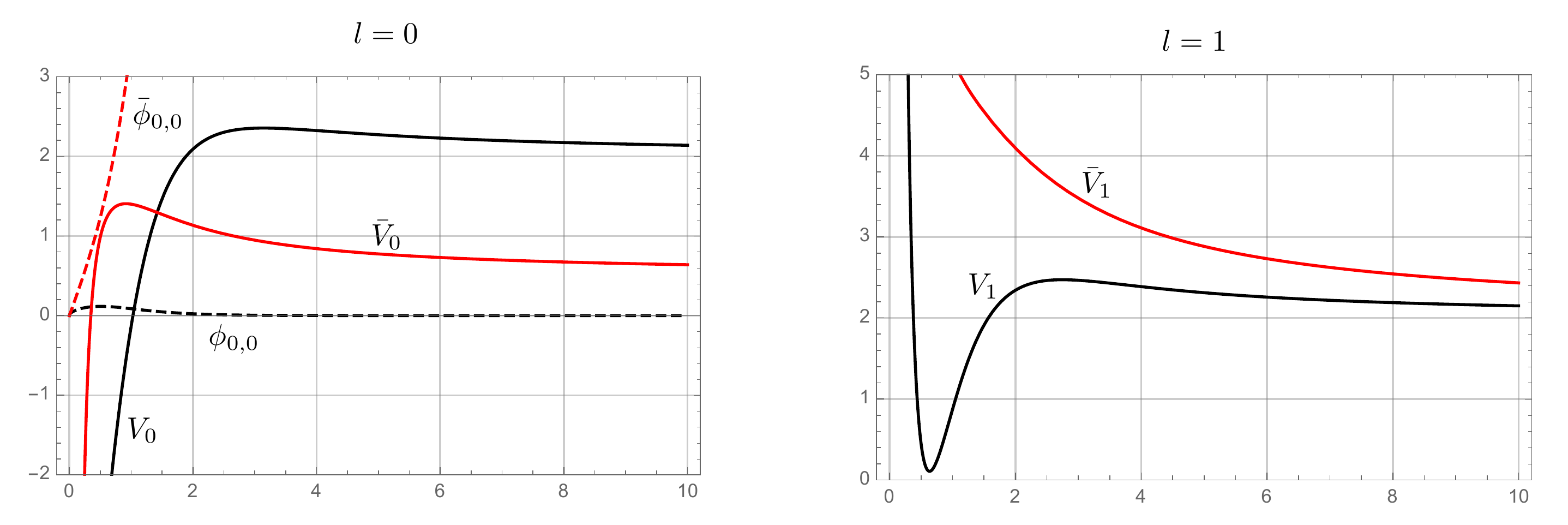}
\caption{The effective potentials for $n=2$. 
See the caption of Fig.~\ref{fig:eff_pot_vor_01} for details.}
\label{fig:eff_pot_vor_02}
\end{center}
\end{figure}

Finally, let us examine the behavior of the potential 
for models with higher power of $\varphi$ as the gauge kinetic 
function $\beta$: namely, we modify the model as
\be
{\cal L}= - \beta^2 {\cal F}_{MN}{\cal F}^{MN} + {\cal L}_{\rm vortex},
\quad \beta = a \varphi^n.
\ee
We plot the effective potentials for $n=2$ case in 
Fig.~\ref{fig:eff_pot_vor_02}.
Compared to the case $n=1$ given in Fig.~\ref{fig:eff_pot_vor_01}, 
the potentials are deeper. 
Therefore, we expect several excited discrete bound states for 
higher $n$.

\subsubsection{Analytic example of the mass spectrum}
\label{sec:analytic_D=6}

\begin{figure}[ht]
\begin{center}
\includegraphics[width=16cm]{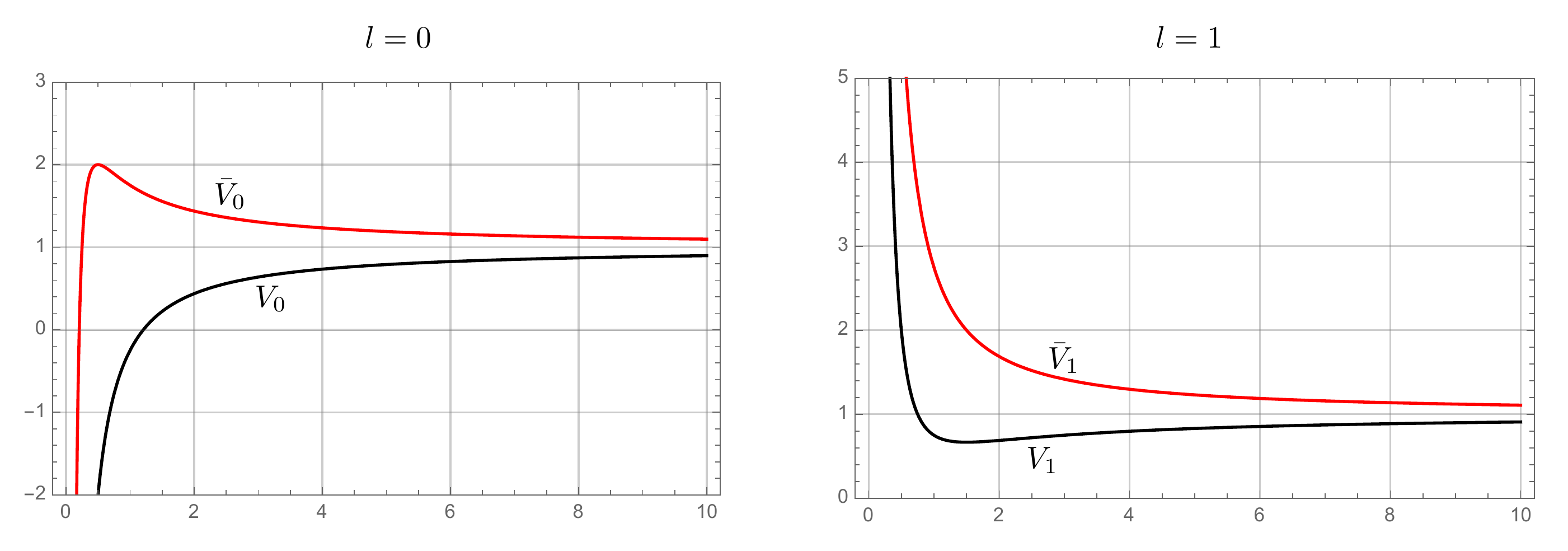}
\caption{The analytic potentials $V_l$ and $\bar V_l$ ($l=0,1$) for $\beta = e^{-\Omega r}$ with $\Omega=1$.
See the caption of Fig.~\ref{fig:eff_pot_vor_01} for details.}
\label{fig:analytic_pot_D6}
\end{center}
\end{figure}

Let us illustrate the results of the previous subsection on a 
concrete example $\beta = e^{-\Omega r}$. 
The relevant eigenvalue equations read 
\be
D^2\ &:&\ \left(-\frac{d^2}{dr^2} + V_l\right) \phi_{n,l} 
= m_{n,l}^2 \phi_{n,l}, \hspace{5mm }V_l 
= \Omega^2-\frac{\Omega}{r}+\frac{l^2-1/4}{r^2}, \\
\bar D^2\ &:&\ \left(-\frac{d^2}{dr^2} +\bar V_l\right) 
\bar \phi_{n,l} = \bar m_{n,l}^2 \bar\phi_{n,l}, \hspace{5mm} 
\bar V_l =   \Omega^2+\frac{\Omega}{r}+\frac{l^2-1/4}{r^2}.
\ee
The difference between $V_l$ and $\bar V_l$ is just 
$\Omega$ and $-\Omega$. 
Fig.~\ref{fig:analytic_pot_D6} shows the potentials which are quite similar to those obtained numerically in Fig.~\ref{fig:eff_pot_vor_01}.
The potential $V_l$ has an extremum at $r^*= (4l^2-1)/(2\Omega)$ with the value
\begin{equation}
V_{l}(r^*) = \Omega^2 \biggl(1-\frac{1}{4l^2-1}\biggr)\,. 
\end{equation}
While $V_0(r)$ is infinitely deep at the origin and $r^*$ is its global maximum, $V_{l\not = 0}$ is unbounded at $r=0$ and has a global minimum at $r^*$. Notice that $V_{l\not=0}(r^*) $ is always lower than the asymptotic value $V_l \to \Omega^2$ as $r\to \infty$. Thus, we expect  for both $V_0$ and $V_{l\not = 0}$ a tower of discrete states. 

\begin{figure}[ht]
\begin{center}
\includegraphics[width=15cm]{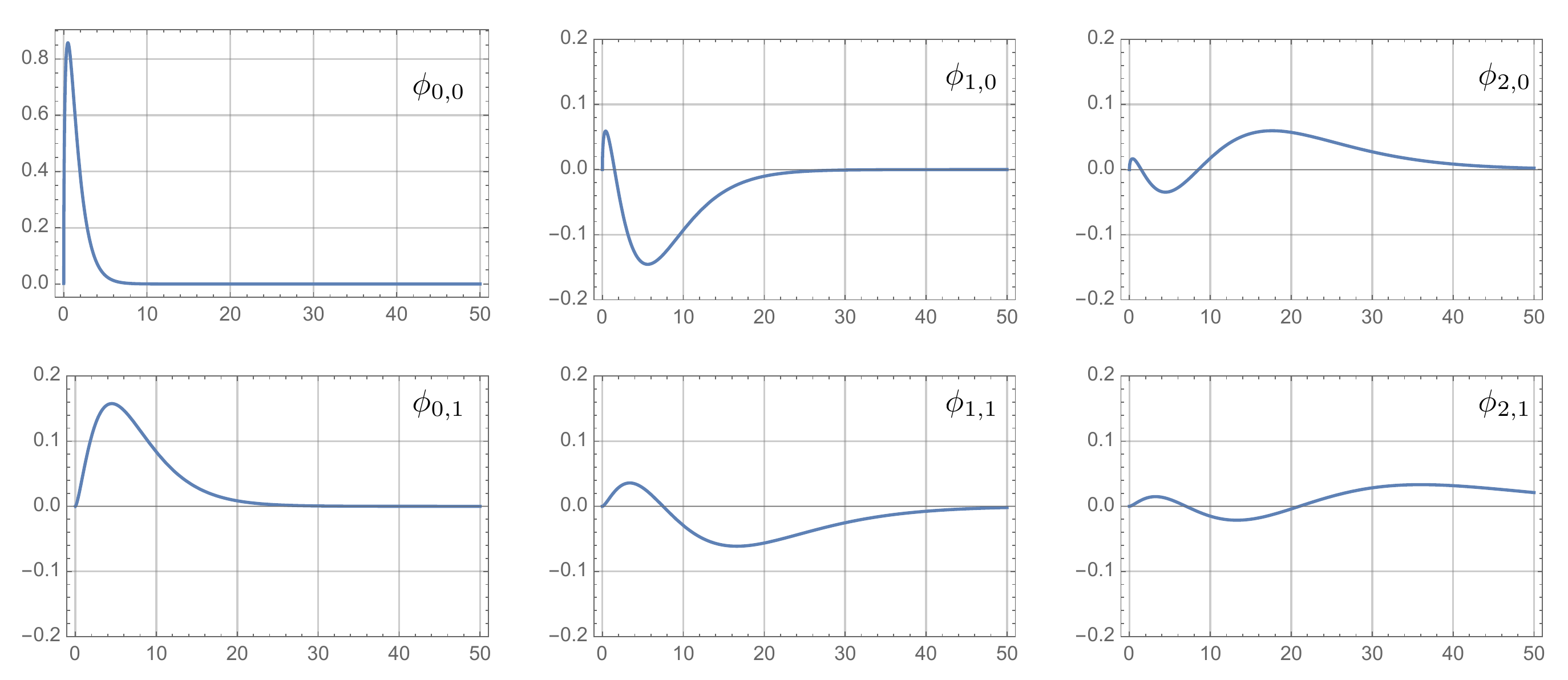}
\caption{The analytic wave functions $\phi_{n,l}$ for $n=0,1,2$ and $l=0,1$  are shown for $\Omega=1$.}
\label{fig:wf_d2_D6}
\end{center}
\end{figure}

Indeed, for each $l = 0,\pm 1,\pm 2,\ldots$
there is an infinite tower of bound states $n=0,1,2,\ldots$ with radial wave functions (up to normalization constant) and eigenvalues given as
\begin{align}
\phi_{n,l}(r) = &\  \exp\biggl(\frac{-\Omega r}{1+2|l|+2n}\biggr)r^{l+\frac{1}{2}} \sum\limits_{k=0}^{n}\frac{(2|l|)!}{(2|l|+k)!}\binom{n}{k}\biggl(\frac{-2 \Omega r}{1+2|l|+2n}\biggr)^k\,, \\
 m_{n,l}^2 = &\ \Omega^2\left(1-\frac{1}{(1+2|l|+2n)^2}\right)\,.
\end{align}
The discrete modes are cumulating at the threshold $m_{n,l} \to \Omega$ as $n\to \infty$, above which there is a continuum labelled by a radial momentum $q$: $m_{q,l} = \sqrt{\Omega^2+q^2}$ with eigenfunctions 
\begin{equation}
\phi_{q,l}(r) = c_1 M\Bigl(-i \Omega/(2q), |l|; 2i q r\Bigr)+c_2 W\Bigl(-i \Omega/(2q), |l|; 2i q r\Bigr)\,,
\end{equation}
where $M\bigl(k,m;z\bigr)$ and $W\bigl(k,m;z\bigr)$ are the Whittaker functions.
For illustration, we show several wave functions in Fig.~\ref{fig:wf_d2_D6}.

\begin{figure}[ht]
\begin{center}
\includegraphics[width=16cm]{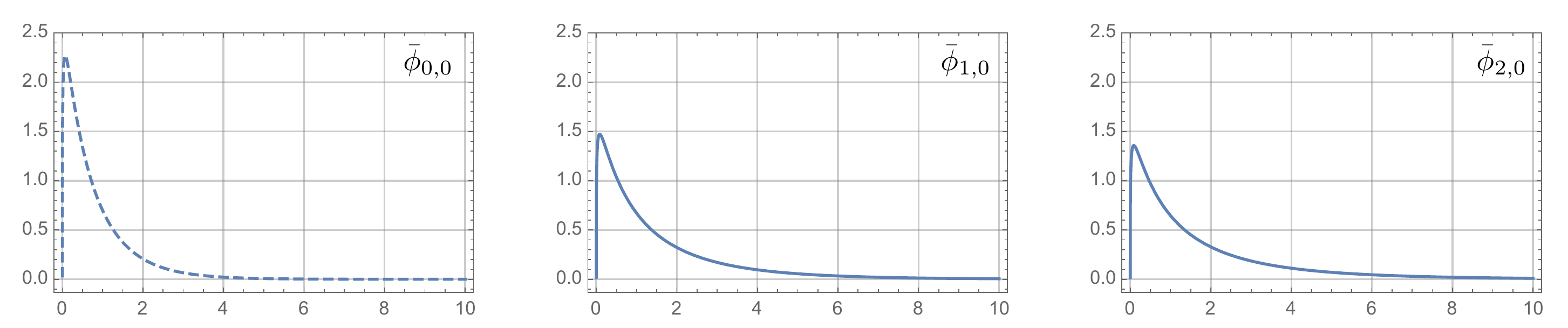}
\caption{The analytic wave functions $\bar \phi_{n,0}$ for $n=0,1,2$ are shown for $\Omega=1$.}
\label{fig:wf_bd2_D6}
\end{center}
\end{figure}

On the other hand, the potential $\bar V_l$ has no minimum for $l\not = 0$ and, in fact, $\bar V_{l\not = 0} > \Omega^2 \equiv \bar V_{l}(\infty)$. Hence, we cannot expect bound states for $V_{l \not = 0}$. However, there is an infinite tower of discrete states for $l=0$ tower with radial wave functions (up to normalization  constant) and eigenvalues given as
\begin{align}
\bar \phi_{n,0}(r) = & \exp\biggl(\frac{\Omega r}{1+2n}\biggr)\sqrt{r} \sum\limits_{k=0}^{n}
\binom{n}{k}(-1)^k E_{k+1}\biggl(\frac{2\Omega r}{1+2n}\biggr)\,, \\
\bar m_{n,0}^2 = &\Omega^2\left(1-\frac{1}{(1+2n)^2}\right)\,.
\end{align}
where $E_n(x) = \int_{1}^{\infty}e^{-x t}t^{-n} dt$ is the Exponential integral. 
Note that, even though the zero mode $\bar\phi_{0,0}$ is normalizable, it does contribute nothing to the physical spectrum as
explained around Eq.~(\ref{eq:L_extra_div_free}).
The eigenfunctions of the continuum part of the spectrum parametrized by a radial momentum $q$ as $\bar m_{q,l} = \sqrt{\Omega^2 +q^2}$ can be expressed in of the Whittaker functions as
 \begin{equation}
\bar\phi_{q,l}(r) = c_1 M\Bigl(i \Omega/(2q), 0; 2i q r\Bigr)+c_2 W\Bigl(i \Omega/(2q), 0; 2i q r\Bigr)\,.
\end{equation}
Fig.~\ref{fig:wf_bd2_D6} shows first few wave functions $\bar\phi_{n,0}$.

Thus, the analytic example here explicitly demonstrates the infinite number of bound states, which
is in contrast to the domain wall case.

\subsection{Spherically symmetric background in $D=7$}
\label{sec:monopole}

In this section, we will investigate physical spectrum for spherically symmetric background 
in $D=7$. Although we do not specify the background solution, we have a codimension three soliton like
a monopole or skyrmion in mind.

\subsubsection{Analysis of a spherically symmetric background}

Let us now investigate the spectrum of gauge fields in arbitrary spherically symmetric background, defined by $\beta(r)$, with $r = \sqrt{x_4^2+x_5^2+x_6^2}$. Following the general discussion of Sec.~\ref{sec:2}, the KK spectrum is determined via equations 
\be
(\p^2 + D^2)A_\mu^{\rm T} = 0,\\
\left[(\p^2 + D^2)\delta_{ab} - D_b^\dagger D_a \right] A_b^{\rm df} = 0\,.
\label{eq:d72}
\ee
where $A_\mu^{\rm T}$ are the transverse four-dimensional gauge fields, i.e. $\partial^\mu A_\mu^{\rm T} = 0$, and $A_a^{\rm df}$ are divergence-free extra-dimensional gauge fields, i.e. $P_{ab}A_b^{\rm df} = 0$.

To fully utilize the spherical symmetry, let us switch to spherical coordinates $x^4 = r \cos\phi\sin\theta$, $x^5 = r \sin\phi\sin\theta$, $x^6 = r \cos\theta$.
The four-dimensional gauge fields satisfy the equation
\begin{equation}
\Bigl(\partial^2 -\Delta +\frac{1}{\beta}\Delta_r \beta\Bigr)A_\mu^{\rm T} = 0\,,
\end{equation}
with $ \Delta_r \equiv \partial_r^2 +\frac{2}{r}\partial_r$ and 
$
\Delta \equiv \Delta_r -\frac{L^2}{r^2},
$
and $L^2 \equiv \partial_\theta^2 +\frac{\cos\theta}{\sin\theta}\partial_\theta+\frac{1}{\sin^2\theta}\partial_\phi^2$.
Let us expand the fields into a common set of four-dimensional zero modes and spherical harmonics
\begin{equation}
A_\mu^{\rm T}(x,r,\theta,\phi) = \sum\limits_n\sum\limits_{l=0}^\infty\sum\limits_{m=-l}^{l}u_\mu^{(n,l,m)}(x)A_{nl}(r)Y_l^m(\theta,\phi)\,,
\end{equation}
where $\partial^\mu u_\mu^{(n,l,m)} = 0$, $\partial^2u_\mu^{(n,l,m)} 
= -\mu_{nl}^2 u_\mu^{(n,l,m)}$ and $L^2 Y_l^m = l(l+1)Y_l^m$. 
The Schr\"odinger equation for the radial wave functions reads
\begin{equation}
\biggl(-\Delta_r +\frac{l(l+1)}{r^2}+\frac{1}{r\beta}\bigl(r\beta\bigr)^{\prime\prime} \biggr)A_{nl} = \mu_{nl}^2 A_{nl}\,.
\label{eq:eigenvalueEq_A}
\end{equation}
As in other examples, there is a unique normalizable zero mode $\mu_{00} = 0$:
\begin{equation}
A_{00}(r) = \beta(r)\,,
\end{equation}
which exists for arbitrary $\beta$.

In order to tackle Eq.~\refer{eq:d72}, we will use the machinery of \emph{vector spherical harmonics} as it is the most convenient tool for separating radial and angular coordinates for vector-valued equations. 
In general, any vector $\vec X \equiv \vec X(r, \theta, \phi)$ can be expanded into the basis of three independent spherical harmonics as
\begin{equation}
\vec X = \sum\limits_{l=0}^{\infty}\sum\limits_{m=-l}^{l}\Bigl(X_{lm}^r(r)\vec Y_l^m+ X_{lm}^{(1)}(r)\vec \Psi_{l}^m+X_{lm}^{(2)}(r)\vec \Phi_l^m\Bigr)\,,
\end{equation}
where $X_{lm}^r, X_{lm}^{(1)}$, $X_{lm}^{(2)}$ are the expansion coefficients and the spherical harmonics $\vec Y_l^m$, $\vec \Psi_l^m$, $\vec \Phi_l^m$ are defined as\footnote{We follow mostly the conventions and notation of \cite{Barrera}.}
\begin{align}
\vec Y_l^m & = \hat r\, Y_l^m\,, \\
\vec \Psi_l^m & = r \vec \nabla\, Y_l^m = \hat \theta\, \partial_\theta Y_l^m +\hat \phi \frac{1}{\sin\theta}\partial_\phi Y_l^m\,, \\
\vec \Phi_l^m & = \vec r\times \vec\nabla\, Y_l^m = \hat \phi\, \partial_\theta Y_l^m -\hat\theta\frac{1}{\sin\theta}\partial_\phi Y_l^m\,,
\end{align}
where $\hat r$, $\hat \theta$, $\hat \phi$ denotes unit vectors in the radial and angular directions.
Vector spherical harmonics have various nice properties. Of particular use for us are the following
\begin{align}
&\Delta \vec Y_l^m = -\frac{l^2+l+2}{r^2}\vec Y_l^m+\frac{2}{r^2}\vec \Psi_l^m\,, \\
&\Delta \vec \Psi_l^m = \frac{l(l+1)}{r^2}\Bigl(2 \vec Y_l^m -\vec \Psi_l^m\Bigr)\,, \\
&\Delta \vec \Phi_l^m = -\frac{l(l+1)}{r^2}\vec \Phi_l^m\,, \\
&\bigl(\vec r \cdot \vec \nabla \bigr) \vec Y_l^m = \bigl(\vec r \cdot \vec \nabla \bigr) \vec \Psi_l^m =\bigl(\vec r \cdot \vec \nabla \bigr) \vec \Phi_l^m = 0\,.
\end{align}
These allow us to establish the key identities for tackling the Eq.~\refer{eq:d72}, namely
\begin{align}
&\Delta \vec X   = \sum\limits_{l=0}^{\infty}\sum\limits_{m=-l}^{l} \biggl[\Bigl(\Delta X_{lm}^r-\frac{l^2+l+2}{r^2}X_{lm}^r+2\frac{l(l+1)}{r^2}X_{lm}^{(1)}\Bigr)\vec Y_l^m  \nonumber \\ 
&\qquad\quad+ \Bigl(\Delta X_{lm}^{(1)}-\frac{l(l+1)}{r^2}X_{lm}^{(1)}+\frac{2}{r^2}X_{lm}^r\Bigr) \vec \Psi_l^m+
\Bigl(\Delta X_{lm}^{(2)}-\frac{l(l+1)}{r^2}X_{lm}^{(2)}\Bigr)\vec \Phi_l^m\biggr]\,, \\
&\bigl(\vec X \cdot \vec \nabla\bigr)\bigl(\hat r f(r)\bigr)  = 
\sum\limits_{l=0}^{\infty}\sum\limits_{m=-l}^{l} \Bigl(X_{lm}^r f^\prime(r) \vec Y_l^m +\frac{1}{r}X_{lm}^{(1)}f(r) \vec \Psi_l^m+\frac{1}{r}X_{lm}^{(2)}f(r)\vec \Phi_l^m\Bigr)\,,
\end{align}
where $(\phantom{T})^{\prime}$ denotes derivative with respect to $r$.

At this point, let us expand the extra-dimensional three-vector 
$\bigl(\vec A\bigr)_a \equiv A_a$ in terms of vector spherical 
harmonics with the four-dimensional effective fields $u^{(n,l,m)}(x)$ 
and $u_2^{(n,l,m)}(x)$ as coefficients  
\begin{align}
\vec A =  
\sum\limits_n\sum\limits_{l=0}^{\infty}\sum\limits_{m=-l}^{l}\biggl[  
u^{(n,l,m)}(x)\Bigl(A_{nl}^r(r)\vec Y_l^m + 
A_{nl}^{(1)}(r)\vec \Psi_l^m\Bigr) +u_2^{(n,l,m)}(x)
A_{nl}^{(2)}(r)\vec \Phi_l^m\biggr]\,,
\end{align}
where 
\begin{equation}
\partial^2 u^{(n,l,m)}(x) = -m_{n,l}^2 u^{(n,l,m)}(x)\,, \hspace{5mm} 
\partial^2 u_2^{(n,l,m)}(x) = -\tilde m_{n,l}^2 u_2^{(n,l,m)}(x)\,.
\end{equation}

Now, the divergence part  $K = D_a^\dagger A_a$ is expanded as
\begin{equation}
K = \sum\limits_n\sum\limits_{l=0}^{\infty}\sum\limits_{m=-l}^{l}u^{(n,l,m)}(x)\biggl(
\partial_r A_{nl}^r +\frac{2}{r}A_{nl}^r-\frac{l(l+1)}{r}A_{nl}^{(1)} + \bigl(\log\beta\bigr)^{\prime} A_{nl}^r\biggr) Y_{l}^m\,.
\end{equation}
In particular, we
 see that $K$ is independent of $A_{nl}^{(2)}$. Thus,  $A_{nl}^{(2)}$ contains only physical degrees of freedom.
Since we are interested in physical degrees of freedom, we set $K=0$. Therefore, for $l=0$
we set $A^{r}_{n0} = 0$ and for $l\neq0$, we eliminate $A^{(1)}_{nl}$ as 
\begin{equation}\label{eq:constr2}
A_{nl}^{(1)} = \frac{r}{l(l+1)}\Bigl(\partial_r A_{nl}^r+\frac{2}{r}A_{nl}^r+ \bigl(\log\beta\bigr)^{\prime}A_{nl}^{r}\Bigr)\,.
\end{equation}
Plugging these into Eq.~\refer{eq:d72}, we find for $l=0$
\begin{align}
\label{eq:e7e2}& \biggl( - \Delta_r +\frac{\beta^{\prime\prime}}
{\beta} \biggr)A_{n0}^{(1)}    = m_{n,0}^2 A_{n0}^{(1)}\,, \\
\label{eq:e7e3}& \biggl( - \Delta_r  + \frac{\beta^{\prime\prime}}
{\beta} \biggr)A_{n0}^{(2)}   = \tilde m_{n,0}^2 A_{n0}^{(2)}\,,
\end{align}
and for $l\neq0$
\begin{align}
\label{eq:e7e4}&\biggl( -\biggl(\partial_r^2+\frac{4}{r}\partial_r\biggr) 
+ \frac{l(l+1)-2}{r^2}+\beta\Big(\frac{1}{\beta}
\Bigr)^{\prime\prime}\biggr)A_{nl}^r  = m_{n,l}^2 A_{nl}^r\,,\\
\label{eq:e7e5}&\biggl( -\Delta_r  + \frac{l(l+1)}{r^2}
+\frac{\beta^{\prime\prime}}
{\beta}\biggr)A_{nl}^{(2)}  = \tilde m_{n,l}^2 A_{nl}^{(2)}\,.
\end{align}

It is clear that $m_{n,0} = \tilde m_{n,0}$ holds.
Notice that for $l=m=0$ vector spherical harmonics takes the values
\begin{equation}
\vec Y_{0}^0 = \frac{\hat r}{\sqrt{4\pi}}\,, \hspace{5mm}
\vec \Psi_{0}^0 = 0\,, \hspace{5mm}
\vec \Phi_0^0 = 0\,.
\end{equation}
As a consequence, we can freely set $A_{00}^{(1,2)}=0$. 
Therefore, we are guaranteed that there are no zero modes for  $A_a^{\rm df}$.

For simplicity, let us introduce 
\begin{equation}
A_{nl} = \frac{B_{nl}}{r},\quad
A_{nl}^r = \frac{B_{nl}^r}{r^2},\quad
A_{n0}^{(1)} = \frac{B_{n0}^{(1)}}{r},\quad
A_{nl}^{(2)} = \frac{B_{nl}^{(2)}}{r}\,.
\end{equation}
Then we finally obtain the following set of one-dimensional 
Schr\"odinger equations from Eqs.~(\ref{eq:eigenvalueEq_A}), 
(\ref{eq:e7e2}), (\ref{eq:e7e4}) and (\ref{eq:e7e5}) as
\begin{align}
&\biggl(-\partial_r^2 +\frac{l(l+1)}{r^2}+\frac{1}{r\beta}\bigl(r\beta\bigr)^{\prime\prime} \biggr)B_{nl}   = \mu_{nl}^2 B_{nl}\,, \\
&\biggl(- \partial_r^2+\frac{l(l+1)}{r^2}+\frac{\beta^{\prime\prime}}{\beta}\biggr)B_{nl}^{(2)}  = \tilde m_{n,l}^2 B_{nl}^{(2)}\,, \\
&\biggl(-\partial_r^2 +\frac{\beta^{\prime\prime}}{\beta}\biggr)B_{n0}^{(1)}   = m_{n,0}^2 B_{n0}^{(1)}\,, \\
&\biggl(-\partial_{r}^2 +\frac{l(l+1)}{r^2}+\beta\Big(\frac{1}{\beta}\Bigr)^{\prime\prime}\biggr)B_{nl}^r  = m_{n,l}^2 B_{nl}^r\,.
\end{align}

\subsubsection{Analytic example of the mass spectrum}

Let us investigate the spectrum on a concrete background $\beta(r) = e^{-\Omega r}$ which substantially simplify
the relevant equations as
\begin{align}
&\biggl(-\partial_r^2 +\frac{l(l+1)}{r^2}+\Omega^2 - \frac{2\Omega}{r} \biggr)B_{nl}   = \mu_{nl}^2 B_{nl}\,, 
\label{eq:mono1}\\
&\biggl(- \partial_r^2+\frac{l(l+1)}{r^2}+\Omega^2 \biggr)B_{nl}^{(2)}  = \tilde m_{n,l}^2 B_{nl}^{(2)}\,, \\
&\biggl(-\partial_r^2 +\Omega^2 \biggr)B_{n0}^{(1)}   = m_{n,0}^2 B_{n0}^{(1)}\,, \\
&\biggl(-\partial_{r}^2 +\frac{l(l+1)}{r^2}+\Omega^2\biggr)B_{nl}^r  = m_{n,l}^2 B_{nl}^r\,.
\end{align}
Eq.~(\ref{eq:mono1}) is nothing but the hydrogen atom with the Coulomb potential. Therefore, $B_{nl}$ for bound states
is the Laguerre polynomials with discrete mass
$ \mu_{n,l}^2 = \ \Omega^2\left(1-\frac{1}{(1+l+n)^2}\right)$.
Continuum modes are labeled by $q$: $\mu_{l}(q) = \sqrt{\Omega^2 +q^2}$.
On the other hand, the Schr\"odinger  potentials for  the fields $B^{(2)}_{nl}$, $B^{(1)}_{n0}$ and $B^r_{nl}$ are constants.
Therefore, no localized modes exist.
Hence, in this example, only four-dimensional components of gauge fields have a discrete tower of localized states, 
while for the extra-dimensional components  there is only a continuum of bulk modes.

\vspace{8mm}
\acknowledgements

F.\ B.\  would like to thank Petr Blaschke for many consultations. M.\ A. and M.\ E. thank to Nobuchika Okada for fruitful discussions.
This work is supported in part by the Albert Einstein Centre for Gravitation and Astrophysics financed by the Czech Science Agency Grant No. 14-37086G (F.\ B.).
This work is also supported in part 
by the Ministry of Education,
Culture, Sports, Science (MEXT)-Supported Program for the Strategic
Research Foundation at Private Universities ``Topological Science''
(Grant No.~S1511006), 
by the Japan Society for the 
Promotion of Science (JSPS) 
Grant-in-Aid for Scientific Research
(KAKENHI) Grant Numbers   
26800119, 16H03984 and 17H06462 (M.\ E.), and 
by the program of Czech Ministry of Education 
Youth and Sports INTEREXCELLENCE Grant number LTT17018 (F. B.).
F.\ B.\ was an international research fellow of the Japan Society 
for the Promotion of Science, and was supported by Grant-in-Aid 
for JSPS Fellows, Grant Number 26004750. 

\appendix

\section{Generalization of the analysis of $A_a$ to $D\ge 7$} 
\label{app:higher_dim}

The operator $H_{ab}$ in Eq.~(\ref{eq:mass_operator}) 
in $N \equiv D-4$ Euclidean dimensions can be factorized as 
\be\label{eq:hdecomp}
H_{ab} = \frac{1}{(N-2)!}\varepsilon_{i_1 \ldots i_{N-2}ad}
\varepsilon_{i_1\ldots i_{N-2}b c}D_d^{\dagger}D_c\,,
\ee
where $\varepsilon_{i_1\dots i_N}$ is a completely anti-symmetric symbol. 
We can rewrite the operator as a product of a $\binom{N}{2}\times N$ 
matrix ${\cal D}$ and its Hermitian conjugate as\footnote{Here the 
Hermitian conjugation acts both on matrix space and on operator space.}
\be\label{eq:hdecomp2}
H = {\cal D}^{\dagger}{\cal D},
\ee 
The row index of ${\cal D}$ spans all $\binom{N}{2}$ inequivalent 
values of the first $N-2$ indices of $\varepsilon_{i_1\ldots i_{N-2}a b}$. 
In this way, the degeneracies are taken care of so that the 
numerical factor in \refer{eq:hdecomp} does not appear in 
\refer{eq:hdecomp2}.
In particular, we have
\begin{align}
{\cal D}^{N=2} &= \bigl(D_5, -D_4\bigr)\,, \\
{\cal D}^{N=3} & = 
\begin{pmatrix}
0 & D_6 & -D_5 \\
-D_6 & 0 & D_4 \\
D_5 & -D_4 & 0
\end{pmatrix}\,, \\
{\cal D}^{N=4} & = 
\begin{pmatrix}
0 & 0 & D_7 & -D_6 \\
0 & -D_7 & 0 & D_5 \\
0 & D_6 & -D_5 & 0 \\
D_7 & 0 & 0 & -D_4 \\
-D_6 & 0 & D_4 & 0 \\
D_5 & -D_4 & 0 & 0
\end{pmatrix}
\end{align}
and so on.

The zero modes of $H$ are annihilated by ${\cal D}$ as
\be
{\cal D} \left(
\begin{array}{c}
\psi_4\\
\vdots \\
\psi_D
\end{array}
\right) = 0. 
\ee
This has an obvious solution with arbitrary function $f$ 
\be
\psi_a = D_a f\,.
\ee
which is valid for arbitrary $\beta$. 

However, this is not suitable  for the divergence free part 
since $(\delta_{ab}-P_{ab}) \psi_b = 0$. 
The examples in subsequent section  
indicate that for certain $\beta$ zero mode in $A_a^{\rm df}$ is possible. 

Beyond these observations it is difficult to establish the
spectrum of $A_{a}^{\rm df}$ in arbitrary dimensions. 
However, the $N=2$ case is analyzed completely as follows. 
Let us introduce a single component ``superpartner" to $H$ as
\begin{equation}
\tilde H = {\cal D}{\cal D}^{\dagger}
 =  \bigl(D_5, -D_4\bigr)\begin{pmatrix}D_5^{\dagger} 
\\ -D_4^{\dagger}\end{pmatrix} = \bar D^2\,.
\label{eq:matrix_H}
\end{equation}
It is well known that the pair of operators  $\{H,\tilde H\}$ 
share the same spectrum expect for the possible zero modes. 
Indeed, if we denote eigenvectors of $H$ as $\psi_\lambda$, i.e. 
$H \psi_\lambda = \lambda \psi$, then ${\cal D}\psi_\lambda$ is 
an eigenvector of $\tilde H$ with exactly the same eigenvalue:
\begin{equation}
\tilde H {\cal D}\psi_\lambda = {\cal D} {\cal D}^{\dagger}
{\cal D}\psi_\lambda = {\cal D} H \psi_\lambda = \lambda 
{\cal D}\psi_\lambda\,.
\end{equation} 
Similarly, denoting the eigenvectors of $\tilde H$ as 
$\tilde \psi_\lambda$, that is $\tilde H \tilde \psi_\lambda 
= \lambda \tilde \psi_\lambda$, we see that 
${\cal D}^{\dagger} \tilde \psi_\lambda$ is an eigenvector of $H$:
\begin{equation}
H {\cal D}^{\dagger}\tilde \psi_\lambda = {\cal D}^{\dagger}
{\cal D}{\cal D}^{\dagger}\tilde \psi_\lambda 
= {\cal D}^{\dagger}\tilde H \tilde\psi_\lambda 
= \lambda {\cal D}^{\dagger}\tilde\psi_\lambda\,.
\end{equation}
Note that the zero mode $\psi_0$ of $H$ does not give a zero 
mode of $\tilde H$.

In the $N=3$ case the superpartner reads
\begin{equation}
\tilde H^{N=3} = 
\begin{pmatrix}
D_5 D_5^{\dagger}+D_6 D_6^{\dagger} & -D_5 D_4^{\dagger} & -D_6 D_4^{\dagger} \\
-D_4 D_5^{\dagger} & D_4 D_4^{\dagger}+D_6 D_6^{\dagger} & -D_6 D_5^{\dagger} \\
-D_4 D_6^{\dagger} & -D_5 D_6^{\dagger} & D_4 D_4^{\dagger}+D_5 D_5^{\dagger}
\end{pmatrix}\,.
\end{equation}
Since the level of complexity in finding the spectrum of this 
operator is about the same as for $H_{ab}^{N=3}$, we gain little 
advantage by switching to the superpartner.  
We will give a concrete analysis specialized for $N=3$ ($D=7$) in 
Sec.~\ref{sec:monopole}.
For $N>3$ cases the situation gets even worse as the superpartner 
is $\binom{N}{2}$-dimensional operator, which is a much larger 
matrix then the original $H_{ab}$. 
We leave as a future problem to derive general results 
about the spectrum of extra-dimensional gauge fields.

\section{The divergence free part in the separable $\beta$}
\label{app:A}

The $\binom{N+1}{2}\times \binom{N+1}{2}$ matrix-valued operator 
$H^{N+1}$ in Eq.~(\ref{eq:matrix_H}) for $N+1$ extra-dimensions 
can be decomposed into the $\binom{N}{2}\times \binom{N}{2}$ 
matrix-valued operator $H^N$ for $N$ extra-dimension and 
one-dimensional subspaces as 
\begin{equation}
H^{N+1} = 
\begin{pmatrix}
H^{N}+D_{N+4}^{\dagger}D_{N+4}\mathbf{1}_N &  - D_{N+4}^{\dagger}\vec D \\
-\vec D^{\dagger} D_{N+4} & D_{N}^2
\end{pmatrix}\,,
\end{equation}
where we denoted an $N$-dimensional vector 
$\vec D^{\dagger} \equiv ( D_4^{\dagger}, \ldots , 
D_{N+3}^{\dagger})$ and $D_N^2 \equiv \vec D^\dagger \cdot \vec{D} 
= D_4^{\dagger}D_4+\ldots +D_{N+3}^{\dagger}D_{N+3}$. 
Let us further decompose the wave-function as 
\begin{equation}
\psi^{(N+1)} = 
\begin{pmatrix}
\psi^{(N)}\\ 
\phi_N
\end{pmatrix}\,.
\end{equation}
Here, $\psi^{(N)}$ denotes an $N$-dimensional vector and $\phi_N$ a scalar. 
Suppressing all labelling of eigenfunctions, the eigenvalue problem 
$H^{N+1}\psi^{(N+1)} = \lambda^{(N+1)}\psi^{(N+1)}$ is rewritten as
\begin{gather}
\label{eq:2eigen1} H^N \psi^{(N)} + D_{N+4}^{\dagger}D_{N+4}
\psi^{(N)} -D_{N+4}^{\dagger}\vec D\, \phi_N 
=  \lambda^{(N+1)}\psi^{(N)}\,, \\
\label{eq:2eigen2} -\vec D^{\dagger} D_{N+4}\psi^{(N)}+D_N^2 \phi_N 
= \lambda^{(N+1)}\phi_N\,.
\end{gather}
Together with these, we also  impose divergence-free condition
\begin{equation}\label{eq:divfreecond}
 \vec D^{\dagger}\psi^{(N)} + D_{N+4}^{\dagger}\phi_N = 0\,.
\end{equation}
Due to this condition, we expect that $\psi^{(N+1)}$ contains 
$N$ independent degrees of freedom, all of which has its own 
tower of eigenmodes.

At this point, let us assume that $\beta$ is separable in at 
least one direction, say, $x^{N+4}$-th.
\begin{equation}
\beta(y_{N+1}) \equiv \beta (y_N)b(x^{N+4})\,.
\end{equation}
Here, we used $y_N \equiv \{x^4, \ldots , x^{N+3}\}$ to denote 
remaining directions. Notice that $b(x^{N+4})$ must be normalizable. 
Since $\beta$ appears in $D_a$ only as $\partial_a \log \beta$, 
the above condition implies that $D_{N+4}$ commutes with all 
other operators. 
Hence, we can also separate the variables in wave-functions as
\begin{equation}
\psi^{(N)}(y_{N+1}) \equiv \psi^{(N)}(y_N) S(x^{N+4})\,, \hspace{5mm}
\phi_N(y_{N+1}) \equiv \phi_{N}(y_N )F(x^{N+4})\,.
\end{equation}

First, let us consider the case where $\psi^{(N)}(y_N)$ is an 
eigenvector of $H^N$ with eigenvalue $\lambda^{(N)}$ and it 
is divergence-free, that is $\vec D^{\dagger}\psi^{(N)} = 0$. 
Eq.~\refer{eq:divfreecond} implies $D_{N+4}^{\dagger}F(x^{N+4}) = 0$. 
However, solving this condition as $F \propto 1/b$ we obtain  
non-normalizable wave-function and, hence, we must set $F=0$. 
Thus, Eq.~\refer{eq:2eigen2} is solved trivially. 
Eq.~\refer{eq:2eigen1} reduces to the eigenproblem for $S(x^{N+4})$ 
in the form
\begin{equation}
D_{N+4}^{\dagger}D_{N+4} S = \bigl(\lambda^{(N+1)}-\lambda^{(N)}\bigr)S\,.
\end{equation}
Denoting the eigenvalues of $D_{N+4}^{\dagger}D_{N+4}$ as 
$\lambda_{N+4}$ we arrive at the solution
\begin{equation}
\psi^{(N+1)}(y_{N+1}) = 
\begin{pmatrix}
\psi^{(N)}(y_N) S(x^{N+4}) \\
0
\end{pmatrix}\,, 
\hspace{5mm}
\lambda^{(N+1)} = \lambda^{(N)}+\lambda_{N+4}\,.
\end{equation}  
In other words, we see that for a separable direction, the Hilbert 
space is a direct product of Hilbert spaces generated by $H^{N}$ 
and $D_{N+4}^{\dagger}D_{N+4}$. Notice that zero mode 
$\lambda^{(N+1)}=0$ can exist only if $\lambda^{(N)} = 0$ does. 
As a consequence, for a fully separable $\beta$ there is no zero 
mode in any number of extra-dimensions, as we can recursively 
apply this argument down to the $N=2$ case, where we establish 
that zero mode does not exists. 

The above solution contains only $N-1$ independent degrees of 
freedom, which are contained in $\psi^{(N)}$. One remaining 
solution can be found by taking $H^{N}\psi^{(N)} =0$. 
In other words, we set
\begin{equation}
\psi^{(N+1)}(y_{N+1}) = 
\begin{pmatrix}
\vec D D_N^{-2} K^{(N)}(y_N) S(x^{N+4})\\
K^{(N)}(y_N) F(x^{N+4})
\end{pmatrix} \,.
\end{equation}
 Here, $K^{(N)}$ stands for divergence part of extra-dimensional 
gauge fields. 
Moreover, the divergence-free condition \refer{eq:divfreecond} 
implies $S = - D_{N+4}^{\dagger}F$. 
Plugging this into Eqs.~\refer{eq:2eigen1}-\refer{eq:2eigen2} 
we ultimately obtain two eigenproblems
\begin{equation}
D_{N+4}D_{N+4}^{\dagger} F = \lambda_{N+4}^{\prime} F\,, \hspace{5mm}
D_N^2 K^{(N)} = \bigl(\lambda^{(N+1)}- \lambda_{N+4}^{\prime}\bigr)K^{(N)}\,.
\end{equation}
As we see, the solution space is again furnished by a direct 
product of solution spaces of two operators. 
One is $D_N^2$, which gives the spectrum to four-dimensional 
gauge fields. 
However, we know that the divergence part $K^{(N)}$ has no zero 
mode and hence $D_N^2 K^{(N)} = \lambda_{4D}^{\prime} K^{(N)}$, 
where prime signals the absence of zero mode in an otherwise 
identical spectrum. The second operator is 
$D_{N+4}D_{N+4}^{\dagger}$, which is just a superpartner to 
$D_{N+4}^{\dagger}D_{N+4}$. Thus, its eigenvalues are 
$\lambda_{N+4}^{\prime}$. 
Putting these observation together, we see that 
$\lambda^{(N+1)} = \lambda_{4D}^{\prime}+\lambda_{N+4}^{\prime}$. 
It is obvious that for this degree of freedom zero mode cannot exits. 

In summary, for a separable $\beta$, we find that the spectrum 
of $N$ independent divergence-free eigenvectors of  $H^{N+1}$ 
can be constructed out of $N-1$ divergence-free eigenvectors 
of $H^{N}$ and $N$-dimensional divergence part $K^{(N)}$ 
combined with eigenfunctions of $D_{N+4}^{\dagger}D_{N+4}$ and 
its superpartner.

\section{$D=5$ with $S^1$ extra-dimension}

Let us give pedagogical derivation of a low energy effective action
from the Abelian gauge theory without matters in the 
$M^{1,3} \times S^1$ spacetime. 
The five-dimensional Lagrangian in the $R_\xi$ gauge is given by 
\be
{\cal L}_\xi = - \frac{1}{4}F_{MN}F^{MN} - \frac{1}{2\xi}f^2 
- \bar c\left(\p_\mu\p^\mu - \xi\p_y^2\right) c,
\ee
with the gauge fixing functional
\be
f = \p^\mu A_\mu - \xi \p_y A_y.
\ee
The gauge-fixing condition eliminates the mixing 
between $A_\mu$ and $A_y$ 
\begin{eqnarray}
{\cal L}_\xi &=& - \frac{1}{4}F_{\mu\nu}F^{\mu\nu} 
- \frac{1}{2}\partial^\nu A_\mu \partial^\nu A_\mu 
- \left(1-\frac{1}{\xi}\right)\partial^\mu A_\mu \partial^\nu A_\nu 
+ \frac{1}{2}\partial_y A_\mu \partial_y A^\mu 
\nonumber \\
&&
+ \frac{1}{2}\partial_\mu A_y \partial_\mu A_y 
- \frac{\xi}{2}(\partial_y A_y)^2 
- \bar c\left(\p_\mu\p^\mu - \xi\p_y^2\right) c. 
\end{eqnarray}
 We expand the gauge field and the ghosts as 
\be
A_M = \sum_n A_M^{(n)}(x) \frac{e^{i\frac{n}{R}y}}{\sqrt{2\pi R}} 
,\quad
c = 
\sum_n c^{(n)}(x) \frac{e^{i\frac{n}{R}y}}{\sqrt{2\pi R}},\quad
\bar c = 
 \sum_n \bar c^{(n)}(x) \frac{e^{i\frac{n}{R}y}}{\sqrt{2\pi R}},
\ee
for $n\in {\mathbb{Z}}$ with $A_M^{(-n)} = A_M^{(n)*}$.
Then we find 
\be
F_{\mu\nu} &=& \frac{F_{\mu\nu}^{(0)}}{\sqrt{2\pi R}} 
+ \sum_{n\neq 0} \frac{e^{i\frac{n}{R}}}{\sqrt{2\pi R}} F_{\mu\nu}^{(n)}
,\\
F_{\mu y} &=& \frac{\p_\mu A_y^{(0)}}{\sqrt{2\pi R}}
+ \sum_{n\neq0}  \frac{e^{i\frac{n}{R}}}{\sqrt{2\pi R}}
\left( \p_\mu A_y^{(n)} - i\mu_n A_\mu\right)
,\\
f &=& \frac{\p^\mu A_\mu^{(0)}}{\sqrt{2\pi R}}
+ \sum_{n\neq0}  \frac{e^{i\frac{n}{R}}}{\sqrt{2\pi R}} 
\left(\p^\mu A_\mu^{(n)}-i\xi \mu_nA_y^{(n)}\right) ,
\ee
with
\be
\mu_n = \frac{n}{R}.
\ee
After integrating the five-dimensional Lagrangian over $y$, 
we obtain a sum of four-dimensional Lagrangians for KK modes as 
\be
{\cal L}_\xi^{\rm eff} = {\cal L}_\xi^{(0)} 
+ \sum_{n=1}^{\infty}  {\cal L}^{(n)}_\xi,
\ee
with 
\be
{\cal L}_\xi^{(n=0)} = - \frac{1}{4}F_{\mu\nu}^{(0)}F^{(0)\mu\nu} 
- \frac{1}{2\xi}\left(\p^\mu A_\mu^{(0)}\right)^2 
- \bar c^{(0)}\p^2 c^{(0)}
+ \frac{1}{2} \p_\mu A_y^{(0)} \p^\mu A_y^{(0)},
\label{app:Lxi_gene0}
\ee
and
\be
{\cal L}_\xi^{(n\neq0)}
&=& 
 A_\mu^{(-n)}\left[(\partial^2+\mu_n^2)\eta^{\mu\nu} 
- \left(1-\frac{1}{\xi}\right)\partial^\mu\partial^\nu\right] A_\mu^{(n)}
\nonumber\\
&&- A_y^{(-n)}(\partial^2+\xi\mu_n^2)A_y^{(n)} 
\nonumber\\
&&
- \bar c^{(-n)} \left(\p^2 + \xi\mu_n^2\right) c^{(n)}
- \bar c^{(n)} \left(\p^2 + \xi\mu_n^2\right) c^{(-n)}.
\label{app:Lxi_n}
\ee
The effective four-dimensional Lagrangian ${\cal L}_\xi^{(0)}$ 
for massless modes is identical to the ordinary $U(1)$ gauge 
theory, a massless scalar field $A_y^{(0)}$ and the ghost field 
associated to the covariant gauge fixing condition. 
Thus we find single massless scalar field as an additional physical 
degree of freedom. 
The effective Lagrangian for massive modes contains massive 
complex vector fields $A_\mu^{(n)}$ with mass $\mu_n$ besides 
ghost fields $c^{(n)}, c^{(-n)}, \bar c^{(n)}, \bar c^{(-n)}$ 
with common gauge-dependent mass squared $\xi\mu_n^2$. 
Their contributions in physical processes cancel each other, and have 
no physical effect. 
This is consistent with the possibility to choose the axial 
gauge where we can eliminate $A_y$ by gauge transformations. 
This gauge choice is possible only as a five-dimensional field, 
and does not exclude possible zero modes, as we find explicitly 
here. 

The presence of the physical massless scalar field $A_y^{(0)}$ 
is a common property of extra-dimensional models with 
{\it compact} extra-dimension. 
The mass gap between massless and massive modes is proportional to $1/R$.
Presence of massless scalar fields is phenomenologically not preferable.
In order to forbid them, it is commonly introduced $Z_2$ parity by 
considering $S^1/Z_2$ orbifold compactification 
\be
A_\mu(x,y) &\to& A_\mu(x,-y) =  A_\mu(x,y),\\
A_y(x,y) &\to& A_y(x,-y) = - A_y(x,y).
\ee
Since the zero mode of $A_y^{(0)}$ is $Z_2$ odd, it is eliminated
from the physical spectrum.


\end{document}